\documentclass{aa}  
\usepackage[usenames,dvipsnames]{color}
\usepackage{comment}

\usepackage[normalem]{ulem}

\usepackage{graphicx}
\usepackage{txfonts}

\usepackage[labelfont=bf]{caption}
\usepackage{natbib}
\usepackage{twoopt}
\usepackage[breaklinks=true]{hyperref} 
\bibpunct{(}{)}{;}{a}{}{,}             
\makeatletter
  \newcommandtwoopt{\citeads}[3][][]{\href{http://adsabs.harvard.edu/abs/#3}%
    {\def\hyper@linkstart##1##2{}%
     \let\hyper@linkend\@empty\citealp[#1][#2]{#3}}}
  \newcommandtwoopt{\citepads}[3][][]{\href{http://adsabs.harvard.edu/abs/#3}%
    {\def\hyper@linkstart##1##2{}%
     \let\hyper@linkend\@empty\citep[#1][#2]{#3}}}
  \newcommandtwoopt{\citetads}[3][][]{\href{http://adsabs.harvard.edu/abs/#3}%
    {\def\hyper@linkstart##1##2{}%
     \let\hyper@linkend\@empty\citet[#1][#2]{#3}}}
  \newcommandtwoopt{\citeyearads}[3][][]%
    {\href{http://adsabs.harvard.edu/abs/#3}
    {\def\hyper@linkstart##1##2{}%
     \let\hyper@linkend\@empty\citeyear[#1][#2]{#3}}}
\makeatother

\usepackage{booktabs}

\usepackage{newtxtext,newtxmath}
\usepackage{anyfontsize}

\usepackage[T1]{fontenc}
\usepackage{placeins}

\DeclareRobustCommand{\VAN}[3]{#2}
\let\VANthebibliography\thebibliography
\def\thebibliography{\DeclareRobustCommand{\VAN}[3]{##3}\VANthebibliography}

\usepackage{tabularx}

\usepackage{graphicx}	
\usepackage{amsmath}	
\usepackage{multirow}
\usepackage{subfig}

\def\be{\begin{equation}} 
\def\ee{\end{equation}} 
\def\ba{\begin{eqnarray}} 
\def\ea{\end{eqnarray}}

\def\msun{{\Msun}}

\def\HH{${\rm {H_2}}$}

\def\gsim{\lower.5ex\hbox{\gtsima}} 
\def\lsim{\lower.5ex\hbox{\ltsima}} \def\gtsima{$\; \buildrel > \over 
\sim \;$} \def\ltsima{$\; \buildrel < \over \sim \;$} \def\prosima{$\; 
\buildrel \propto \over \sim \;$} \def\gsim{\lower.5ex\hbox{\gtsima}} 
\def\lsim{\lower.5ex\hbox{\ltsima}} 
\def\simgt{\lower.5ex\hbox{\gtsima}} 
\def\simlt{\lower.5ex\hbox{\ltsima}} 
\def\simpr{\lower.5ex\hbox{\prosima}}   
  
 \def\gtsima{$\; \buildrel > \over \sim \;$} 
\def\ltsima{$\; \buildrel < \over \sim \;$} 
\def\gsim{\lower.5ex\hbox{\gtsima}} 
\def\lsim{\lower.5ex\hbox{\ltsima}} 
\def\simgt{\lower.5ex\hbox{\gtsima}} 
\def\simlt{\lower.5ex\hbox{\ltsima}} 
\def\simpr{\lower.5ex\hbox{\prosima}}

\def\msun{\,{\rm \Msun}}

\def\E3{{\cal E}_{\rm g}^{III}}

\def\r12{r_{1/2}} 
\def\x12{x_{1/2}} 
\def\v12{v_{1/2}}

\newcommand\code[1]{\textsc{\MakeLowercase{#1}}}

\def\nh2{n_{\rm H2}}
\def\fh2{f_{\rm H2}}

\def\angstrom{\textrm{A\kern -1.3ex\raisebox{0.6ex}{$^\circ$}}}

\def\mum{\mu{\rm m}}
\def\msun{{\rm M}_{\odot}}
\def\zsun{{\rm Z}_{\odot}}
\def\lsun{{\rm L}_{\odot}}

\def\AGNfiducial{\emph{AGN\_fid}}

\usepackage{xcolor}

\defcitealias{Chakraborty_2023}{C23}
\defcitealias{Valentini:2021}{V21}

\begin{document} 
   \title{Multi-wavelength properties of $z\gtrsim 6$ LISA detectable events}

   \subtitle{}

    \author{Srija Chakraborty
        \inst{1}\fnmsep\thanks{email: srija.chakraborty@sns.it}
          \and
          Simona Gallerani\inst{1}
          \and
          Fabio Di Mascia\inst{1} 
          \and
          Tommaso Zana\inst{2}
          \and
          Milena Valentini\inst{3,4,5,6} 
          \and
          Stefano Carniani\inst{1}
          \and
          Fabio Vito\inst{7}
          \and
          Maulik Bhatt\inst{1}
          }

    \institute{Scuola Normale Superiore, Piazza dei Cavalieri 7, 56126 Pisa PI, Italy
    \and
    Dipartimento di Fisica, Sapienza, Universita di Roma, Piazzale Aldo Moro 5, 00185 Roma, Italy
    \and
    Dipartimento di Fisica dell'Universita di Trieste, Sez. di Astronomia, via Tiepolo 11, I-34131 Trieste, Italy
    \and
    INAF, Osservatorio Astronomico di Trieste, via Tiepolo 11, I-34131, Trieste, Italy
    \and
    INFN, Instituto Nazionale di Fisica Nucleare, Via Valerio 2, I-34127, Trieste, Italy
    \and
    ICSC - Italian Research Center on High Performance Computing, Big Data and Quantum Computing, via Magnanelli 2, 40033, Casalecchio di Reno, Italy
    \and
    INAF - Osservatorio Astronomico di Bologna, via Piero Gobetti 93/3, I-40129 Bologna, Italy}

   \date{Received XX/XX/2024; accepted XX/XX/XXXX}

\abstract
{We investigate the intrinsic and observational properties of $z\gtrsim 6$ galaxies that host the coalescence of massive ($M_{\rm BH}\sim 10^{5-6}~\rm M_{\odot}$) black holes (MBHs) giving rise to gravitational waves (GWs) detectable with the Laser Interferometer Space Antenna (LISA).  We adopt a zoom-in cosmological hydrodynamical simulation of galaxy formation and black hole (BH) co-evolution, based on the GADGET-3 code, zoomed-in on a $M_h \sim 10^{12}~\rm M_{\odot}$ dark matter halo at $z = 6$, which hosts a fast accreting ($\dot{M}\sim 35~\rm M_{\odot}~yr^{-1}$) super-massive black hole (SMBH, $M_{\rm BH}\sim 10^{9}~\rm M_{\odot}$) and a 
 star-forming galaxy (SFR $\sim 100~\rm M_{\odot}~yr^{-1}$).
 Following the SMBH formation backward in time, we identify the merging events that concurred to its formation and we pick up the ones that are detectable with LISA. Among these LISA detectable events (LDEs), we select those that, based on their intrinsic properties ($\dot{M}$, SFR, gas metallicity, and dust mass), are expected to be bright in one or more electromagnetic (EM) bands, e.g. rest-frame X-ray, ultra-violet (UV) and far-infrared (FIR). We further restrict our selection to those LDEs that, after considering the effect of delay due to dynamical friction in the MBH coalescence, are still occurring at $z\gtrsim 6$. We find that $\sim$20-30\% of the LDEs and their host galaxies are also detectable with EM telescopes. We post-process these events with dust radiative transfer calculations to make accurate predictions about their spectral energy distributions (SEDs) and continuum maps in the JWST to ALMA wavelength range. We compare the spectra arising from galaxies hosting the merging MBHs with those arising from AGN powered by single accreting BHs.
 We find that it will be impossible to identify an LDE from the continuum SEDs because of the absence of specific imprints from the merging MBHs. Finally, we compute the profile of the H$_{\rm \alpha}$ line arising from LDEs, taking into account both the contribution from their star-forming regions and the accreting MBHs. We find that the presence of two accreting MBHs would be difficult to infer even if both MBHs accrete at super-Eddington rates ($\lambda_{\rm EDD}\sim 5-10$).

 We conclude that the combined detection of GW and EM signals from $z\gtrsim 6$ MBHs is challenging (if not impossible) not only because of the poor sky-localization ($\sim$10~$\rm deg^2$) provided by LISA, but also because the loudest GW emitters ($M_{\rm BH}\sim 10^{5-6}~\rm M_{\odot}$) are not massive enough to leave significant signatures (e.g. extended wings) in the emission lines arising from the broad line region.}


   \maketitle



\section{Introduction} \label{sec:Introduction}

Follow-up observations of $z \sim 6-7.5$ quasars, the brightest (bolometric luminosities 
$L_{\rm bol} \gtrsim 10^{46}$~erg s$^{-1}$) and the most distant active galactic nuclei (AGN) discovered so far, have shown that these sources are powered by super-massive black holes  \citep[SMBHs, $10^8-10^{10}\rm M_{\odot}$, see a recent review by][and references therein]{fan2023}. The existence of these SMBHs is puzzling theoretical models of BH formation and evolution that are striving to understand both the origin and mass of SMBH seeds, and their ability to grow fast enough to assemble SMBHs in less than 1 Gyr \citep[the age of the Universe at $z\gtrsim 6$; e.g.][]{Volonteri03, latif2016,valiante2017,volonteri2021}. 

The most recent \textit{James Webb Space Telescope} (JWST) data has revealed the presence of accreting MBHs ($10^6-10^{8}\rm M_{\odot}$) in galaxies up to $z\sim 10-11$ \citep[e.g.][]{bosman2023,greene2023,goulding2023,furtak2023,kokorev2023,larson2023,maiolino2023a,kovacs2024},
providing an unprecedented testing ground for theoretical models \citep[e.g.][]{jeon2024,trinca2023,pacucci2023, schneider2023,bhatt2024}. In particular, JWST data has revealed the presence of dual AGN at $z\sim 6-7$ \citep{ubler2023,matsuoka2024}, powered by pairs of accreting MBHs. Since dual AGNs are considered to be the precursors of merging MBHs \citep{saeedzadeh2024}, these findings have been interpreted as the evidence that merging of MBHs in the distant Universe is common. On one hand, the existence of MBH merging is expected from theoretical models; on the other hand, the actual number of MBH mergers depends on several properties of the MBHs and their host galaxies.

According to the hierarchical structure formation paradigm, large galaxies are assembled through the merging of smaller galaxies \citep{PressSchechter_1974}. Since MBHs are nested in the nuclei of their host galaxies \citep{Lynden_Bell,Wise2019Natur.566...85W}, galaxy mergers can lead to the formation of MBH pairs \citep{Yu_2002,Volonteri_2021}.
  
Depending on the mass ratio of the MBH pair, on the initial separation between the MBHs, and on the physical properties of their host galaxies, the dynamical friction exerted by the surrounding gas and stars on the MBHs \citep[e.g.][]{begelman1980, Armitage_2002,Sesana_2015}  may reduce their initial separation and modify their dynamics \citep[see the recent paper by][and references therein]{damiano2024}, eventually leading to their coalescence and the subsequent emission of gravitational waves (GWs). 

The European Space Agency’s Science Programme Committee has recently accepted the challenge of detecting and studying GWs from space with the approval of the Laser Interferometer Space Antenna (LISA) mission. Given its sensitivity and frequency coverage ($10^{-4} - 10^{-1}$ Hz), LISA is expected to detect GWs from various low frequency sources, including GWs arising from MBH ($10^4-10^{7}\rm M_{\odot}$) mergers at high redshift ($z\sim 6-10$) \citep{Amaro_Seoane_2023}
, similarly to what can be done with proposed missions such as TianQin \citep{luo2016} and Taiji \citep{ruan2020}. The thrilling results obtained by the NANOGrav Pulsar Timing Array (PTA), Parkes PTA \citep{Manchester_2013}, International PTA \citep{Verbiest_2016, Antoniadis_2022}, Indian PTA \citep{Tarafdar_2022}, Chinese PTA \citep{Xu_2023}, and the European PTA \citep{EPTA2023, agazie2023}, that found a 3$\sigma$ evidence for a stochastic GW background, likely originating from the mergers of MBHs, is further encouraging the studies of merging MBHs.

Furthermore, the joint detection of GWs and electromagnetic (EM) signals from the host galaxies of the merging MBHs would provide several exciting opportunities: i) to obtain independent constraints on cosmological parameters, by comparing the luminosity distance directly observable by GWs with the redshift inferred from the EM emission \citep[e.g.][]{schutz1986,abbott2018}; ii) to study BH binaries properties (e.g. masses, orbital parameters, Eddington ratios) and characterize their environment \citep[e.g.][]{bogdanovic22}; iii) to clarify open issues related to the formation of SMBHs at high redshift \citep[$z\gtrsim 6$; e.g. ][]{Sesana_2004,lops2023}. Nevertheless, the detection of EM signatures from LISA detectable events (LDEs) is extremely challenging because of the poor sky-localization ($\sim$10~$\rm deg^2$) provided by LISA \citep[e.g.][]{Mangiagli_2020, lops2023, Chakraborty_2023}. 
For these reasons, any hint about the properties of the host galaxies hosting LDEs is strikingly useful. 

The detectability of EM signatures from LDEs has been investigated both through semi-analytical models and numerical simulations. For what concerns $z\lesssim3$, \citet{Izquierdo_Villalba_2023} used semi-analytical model to explore host galaxy properties of LISA MBHBs at this redshift range and showed that the hosts display properties of optically dim, gas-rich, star-forming, disk dominated, low mass galaxies. However, it is challenging to distinguish LISA host galaxy from galaxies hosting singular BHs with similar mass.  \citet{Valiante_2020} adopted the GAMETE semi-analytical model finding that mergers involving heavy seeds ($\sim 10^5-10^7 \rm M_{\odot}$) are detectable by Athena (JWST) up to $z\sim 16$ ($z\sim 13$). By exploiting a semi-analytical galaxy formation and evolution model, \citet{mangliali22} showed that LDEs are expected to be associated with faint EM emission, challenging the observational capabilities of future telescopes, and possibly providing few ($\sim 7-20$) multi-messenger detections of merging MBHs.  The analysis of the OBELISK simulations \citep{Dong_Paez_2023} suggests that at $z>3.5$ the signal arising from merging BHs is fainter than their star-forming host galaxies, thus being hardly detectable in the rest frame UV. Conversely, the X-ray emission of 5-15\% of the merging BHs analysed is sufficiently bright to
be detected with sensitive X-ray instruments. Numerical simulations have been also used to investigate whether morphological \citep[e.g.][]{degraf2020morphological,Bardati23} and/or spectral signatures \citep{Bardati24} are associated with MBH mergers, finding that it is possible to statistically identify their host galaxies, with an accuracy that increases with the chirp mass and the mass ratio.

In this work, we explore the results presented in \citet[][hereafter \citetalias{Chakraborty_2023}]{Chakraborty_2023}, based on the zoom-in cosmological hydro-dynamical simulations developed by \citet[][hereafter \citetalias{Valentini:2021}]{Valentini:2021}. Here, we focus on MBH pairs that, according to the prescriptions of the simulations, are expected to merge at $z\gtrsim 6$. We then carry out the following steps: (i) we characterise their intrinsic properties (e.g. the star formation rate of the host galaxy and the accretion rates of the MBHs), (ii) we compute their observable properties (e.g. the UV, FIR, and X-ray emissions), and (iii) we check whether they are detectable with current and upcoming EM telescopes (e.g. ALMA, JWST, Chandra, Lynx). In particular, for what concerns (ii), we post-process the hydrodynamical simulations with radiative transfer calculations by exploiting the \code{SKIRT} code. The paper is organised as follows: in Sec. \ref{sim} we present the simulations adopted and post-processed with radiative transfer calculations; in Sec. \ref{selecRT}, we select the LDEs that are the most promising for a possible EM detection; in Sec. \ref{RT} we present our results, including the outcomes of RT simulations; in Sec. \ref{halpha}, we compute the shape of the H$\alpha$ line from MBH pairs; we finally summarise and discuss the main findings of our work in Sec. \ref{dis}.

\section{Numerical Models}\label{sim}

In this Section, we describe the main properties of the cosmological hydrodynamical simulations analysed in this work as well as the radiative transfer code implemented in post-processing. We select the \AGNfiducial{} run of the suite presented by \citetalias{Valentini:2021}, previously analysed by \citetalias{Chakraborty_2023} to compute the GW properties of the merging MBHs predicted by simulations.

\subsection{ Hydrodynamical simulations}
The simulations are performed with the TreePM (particle mesh) + SPH (Smoothed Particles Hydrodynamics) code \code{GADGET-3}, an evolution of the public \code{GADGET-2} code \citep{Springel2005Gadget}, 
and follow the evolution of a halo whose mass is $\sim 10^{12}$~M$_{\odot}$ at $z=6$.

In particular, we consider the \AGNfiducial{} run from \citetalias{Valentini:2021}, featuring star formation, stellar feedback, and thermal AGN feedback, among other physical processes.
 
We summarise in the following sections the main features of the simulations that are relevant to the present study, while we refer to the aforementioned papers for details.

\subsubsection{Initial conditions and resolution}
\label{subsec:Valentini_ICs}

The code \code{MUSIC}\footnote{\code{MUSIC}–Multiscale Initial Conditions for Cosmological Simulations: \url{https://bitbucket.org/ohahn/music}.} \citep{Hahn_2011} is used to generate the initial conditions of this simulation\footnote{A $\Lambda$CDM cosmology is assumed with the following parameters \citep{Planck2016}: ${\Omega_{\rm M,0}= 0.3089}$, ${\Omega_{\rm \Lambda,0}= 0.6911}$, ${\Omega_{\rm B,0}= 0.0486}$, ${H_0 = 67.74~\rm{km~s}^{-1}~{\rm Mpc}^{-1}}$.}. 
First, a dark matter (DM)-only simulation is run from $z=100$ to $z=6$, DM particles having a mass of $9.4\times 10^8~\msun$ in a comoving volume of $(148~{\rm Mpc})^3$. A halo as massive as $M_{\rm halo}=1.12 \times 10^{12}~\msun$ at $z=6$ is selected for the zoom-in, full-physics simulation. 
In the zoom-in region, the highest resolution particles have a mass $M_{\rm DM}=1.55\times 10^6~\msun$ and $M_{\rm gas}=2.89 \times 10^5~\msun$. The gravitational softening lengths employed are $\epsilon_{\rm DM}=0.72$~ckpc\footnote{We use the following convention when indicating distances: a letter \emph{c} before the corresponding unit refers to \emph{comoving} distances (e.g. ckpc), while the letter \emph{p} refers to \emph{physical} units (e.g. pkpc). When not explicitly stated, we are referring to physical distances.} and $\epsilon_{\rm bar}=0.41$~ckpc for DM and baryon particles respectively.

\subsubsection{Sub-resolution physics}\label{BH-model}
\begin{enumerate}
\item[$\bullet$]\ Cooling, star formation and stellar feedback{}:\\ 
A multiphase interstellar medium (ISM) described by the MUlti Phase Particle Integrator (MUPPI) sub-resolution model \citep{Murante2010, Murante2015, Valentini2017, Valentini2018, Valentini2019} is used. It features metal-lines cooling, an \HH-based star formation criterion, thermal and kinetic stellar feedback, the presence of a UV background, and a model for stellar chemical evolution \citep{Tornatore_2007}. 

\item[$\bullet$]\ Black holes seeding and merging{}:\\
Black holes are treated as collisionless sink particles with a seed mass of $M_{\rm BH, seed} = 1.48 \times 10^5~\msun$ seeded in DM halos when they first exceed a mass of $M_{\rm DM, seed}=1.48 \times 10^9~\msun$. This seeding prescription is meant to capture the results of the so-called direct collapse BH scenario which predicts SMBH seeds of mass $M_{\rm BH, seed}\sim 10^4-10^6~\msun$ \citep{ferrara2014,haehnelt1993,mayer2019}.
Two BHs merge when the following conditions are both satisfied: (i) their relative distance becomes smaller than twice the BH gravitational softening length; (ii) their relative velocity is lower than the sound speed of the local ISM. The merger is instantaneous, i.e. no time delays between binary formation and coalescence are considered. The final BH as the product of the collision occupies the position of the more massive BH which underwent the merger.
BH repositioning or \emph{pinning} is also implemented, to prevent BHs from wandering from the centre of the halo in which they reside: at each time-step BHs are shifted towards the position of minimum gravitational potential within their softening length instantaneously \citep[as also done in e.g.][]{Booth_2009, Schaye2015MNRAS, Weinberger2017, Pillepich2018}.

\item[$\bullet$]\ BH accretion{}:\\ 
Along with BH-BH mergers, BHs are also allowed to grow via gas accretion from the surroundings. Gas accretion is described by the classical Bondi-Hoyle-Lyttleton accretion solution \citep{Hoyle1939, Bondi_Hoyle, Bondi52, Edgar_2004}:

\begin{equation}  
\label{eq-Mdot-Bondi} 
\dot{M}_{\rm Bondi} = \frac{4 \pi G^2 M_{\rm BH}^2 \rho}{ \left(c_{\rm s}^2 + v^2\right) ^ {3/2}} , 
\end{equation}

where $G$ is the gravitational constant, $\rho$ is the gas density, $c_{\rm s}$ is its sound speed, and $v$ is the velocity of the BH relative to the gas. These quantities are evaluated by averaging over the SPH gas particles within the BH smoothing length, with kernel-weighted contributions. 
Eq. \ref{eq-Mdot-Bondi} is used to estimate the contribution to the accretion rate from the cold and hot phase of the ISM separately \citep{Steinborn2015, Valentini2020}. Accretion from the cold gas is reduced by taking into account its angular momentum \citep[see][for details]{Valentini2020}. 
The BH accretion rate is capped to the Eddington accretion rate.

\item[$\bullet$]\ AGN feedback{}: \\
A fraction of the accreted rest-mass energy is radiated away with a radiative efficiency $\epsilon_{\rm r}$, thereby providing a bolometric luminosity for a BH equal to:
\begin{equation} 
\label{eq-Lr-BH} 
L_{\rm bol} = \epsilon_{\rm r} \dot{M}_{\rm BH}c^2,
\end{equation}
where $c$ is the speed of light and $\epsilon_{\rm r} = 0.03$  \footnote{The choice of the radiative efficiency value agrees with the minimum accretion efficiency of a non-spinning BH surrounded by an accretion disc \citep{Shakura_Sunyaev} and is also compatible with the results by \citet{Sadowski2017} and \citet{Trakhtenbrot_2017}.}. 
A fraction $\epsilon_{\rm f} = 10^{-4}$ (\citetalias{Valentini:2021}) of the radiated luminosity $L_{\rm bol}$ is coupled thermally and isotropically to the gas surrounding the BH. This AGN feedback energy is distributed to the hot and cold phases of multiphase gas particles within the BH smoothing volume \citep{Valentini2020}.
\end{enumerate}

\subsection{Galaxy-merging MBHs association} 
\label{G-BH}

From the cosmological hydrodynamical simulations, we identify merger events as described in \citetalias{Chakraborty_2023}.

To associate a host galaxy to a merger event, we follow the procedure described in \citet{Zana2022} (see also Sec. 4.1 in \citetalias{Chakraborty_2023}). We assign each merger event to the galaxy whose center of mass is closest to the position of the most massive BH. Hereafter, we call "primary BH" (BH$_p$) and "secondary BH" (BH$_s$), the most and least massive BH, respectively. 

We underline that, as mentioned in Sec. \ref{BH-model}, in \citetalias{Valentini:2021} simulations, the merger between two MBHs occurs instantaneously. However, the actual coalescence of two MBHs depends on their interaction with gas and stars in their surroundings which allows the MBHs to lose energy 
and to spiral inwards gradually \citep{Chandrasekhar1943, Ostriker_1999}. This process delays the MBH merger with respect to what is assumed in the adopted simulations. As in \citetalias{Chakraborty_2023}, we correct in post-processing the coalescing time of MBH mergers including a time delay due to dynamical friction from the surrounding stars. In this case, the dynamical friction timescale is computed as \citep{Krolik_2019,Volonteri_2020}:
\begin{equation}
t_{\rm df}=0.67\, {\rm Gyr} \left(\frac{a}{4\, {\rm kpc}}\right)^2\left(\frac{\sigma}{100 \, {\rm km\, s^{-1}}}\right)\left(\frac{M_{\rm BH_s}}{10^8 \,M_{\odot}}\right)^{-1}\frac{1}{\Lambda},
\label{eq:tdf}
\end{equation}
where $a$ is the distance of the BH$_p$\footnote{The distance $a$ is an output of the simulations, and we report physical kpc as the distance between galaxy and MBH in Tab.~\ref{intrinsicprops} and \ref{observable}, as derived from the simulations.} from the galaxy centre, $\sigma$ is the stellar velocity dispersion, and $\Lambda$ is given by:
\begin{equation}
\Lambda=\ln(1+M_{*}/M_{\rm BH_s}), 
\label{eq:tlam}
\end{equation}
 where $M_*$ and $M_{\rm BH_s}$ denote the stellar mass of the host galaxy and the mass of the BH$_s$, respectively.

\subsection{Radiative Transfer calculations} 
\label{RTsim}

We perform Radiative Transfer (RT) calculations including dust with the code \code{SKIRT}\footnote{Version 8, \url{http://www.skirt.ugent.be}.}, which is a Monte-Carlo radiative transfer solver, designed to model radiation fields in dusty media, accounting for dust grains scattering and absorption, and their ensuing re-emission in the IR \citep[e.g.][]{Baes:2003, Camps_2015}. We use a similar numerical setup to the one used in \citealt{DiMascia:2021}, which we summarise below.

For the RT calculations, we select a cubic region with a side length of $40$~kpc, centered on the centre of mass of the most-massive halo, where the galaxies associated with the merging events reside.
This region is then post-processed in \code{SKIRT} by using a computational box of the same size. The RT simulation requires two main ingredients: a source component (e.g. stars, AGN), which determines the radiation field before accounting for dust attenuation, and a dust component, which absorbs and scatters the radiation, and then thermally re-emits photons, altering the radiation field. We describe how these two components are imported from the hydrodynamic simulation in \code{SKIRT} in the following subsections.

\subsubsection{Dust properties} \label{sec:dust_properties}
Dust is distributed in the \code{SKIRT} computational domain by assuming a linear scaling with the gas metallicity\footnote{Throughout this paper the gas metallicity is expressed in solar units, using ${\zsun=0.013}$ as a reference value \citep{Asplund:2009}.} \citep[e.g.][]{Draine_2007}, according to a \emph{dust-to-metal ratio} $f_{\rm d}$ that quantifies the mass fraction of metals locked into dust:
\begin{equation} \label{eq:fd}
    f_{\rm d} = M_{\rm dust} / M_Z,
\end{equation}
where $M_{\rm dust}$ is the dust mass and $M_Z$ is the total mass of all the metals in each gas particle in the hydrodynamical simulation.
Gas particles hotter than $10^6$~K are assumed to be dust-free because of thermal sputtering \citep[e.g.][]{Draine:1979}. We adopt the value of $f_{\rm d}=0.1$, which is consistent with RT simulations reproducing the emission of high-redshift galaxies \citep[e.g.][]{Behrens:2018}.

The dust distribution derived from the gas particles is then distributed into a dust grid in \code{SKIRT}. We adopt an octree grid with grid with nine levels of refinement, corresponding to a maximum spatial resolution of $\approx 80$~pc, which is comparable to the resolution of the hydrodynamical simulations at the redshifts of the events. 
We assume the dust composition to be the one that reproduces the extinction curve of the Small Magellanic Cloud (SMC), by using the results by \citealt{Weingartner:2001}.
We take into account dust self-absorption in our calculations. We instead neglect non-local thermal equilibrium (NLTE) corrections to dust emission. We do not include heating from CMB radiation. 

\subsubsection{Stellar and AGN radiation}

Stellar and BH particles are treated as point sources of radiation. We describe the stellar emission by using the family of stellar synthesis models by \citealt{Bruzual:2003}. For the black holes, we adopt the AGN fiducial spectral energy distribution (SED) introduced in \citealt{DiMascia:2021}, which can be written as a composite power-law:
\begin{equation}\label{AGN_SED_eq2}
     L_\lambda = c_i \ \left(\frac{\lambda}{\mu{\rm m}}\right)^{\alpha_i} \ \left(\frac{L_{\rm bol}}{\lsun}\right) \ \lsun \ {\mum}^{-1},
\end{equation}
where $i$ labels the bands in which we decompose the spectra and the coefficients $c_i$ are determined by imposing the continuity of the function based on the slopes $\alpha_i$ \citep[see Table 2 in][for specific values of the coefficients]{DiMascia:2021}.
The SED is then normalized according to the bolometric luminosity of the black hole, as expressed by eq. \ref{eq-Lr-BH}.

The radiation field is sampled by using a grid composed of $200$ logarithmically spaced bins, covering the \emph{rest-frame} wavelength range ${[0.1-10^3]~\mum}$. A total of $10^6$ photon packets per wavelength bin is launched from each source\footnote{We verified that this number is high enough to achieve the convergence of the results of the RT calculations.}.

\begin{figure*}
        \centering
        
          \includegraphics[width=1.0\textwidth]
          {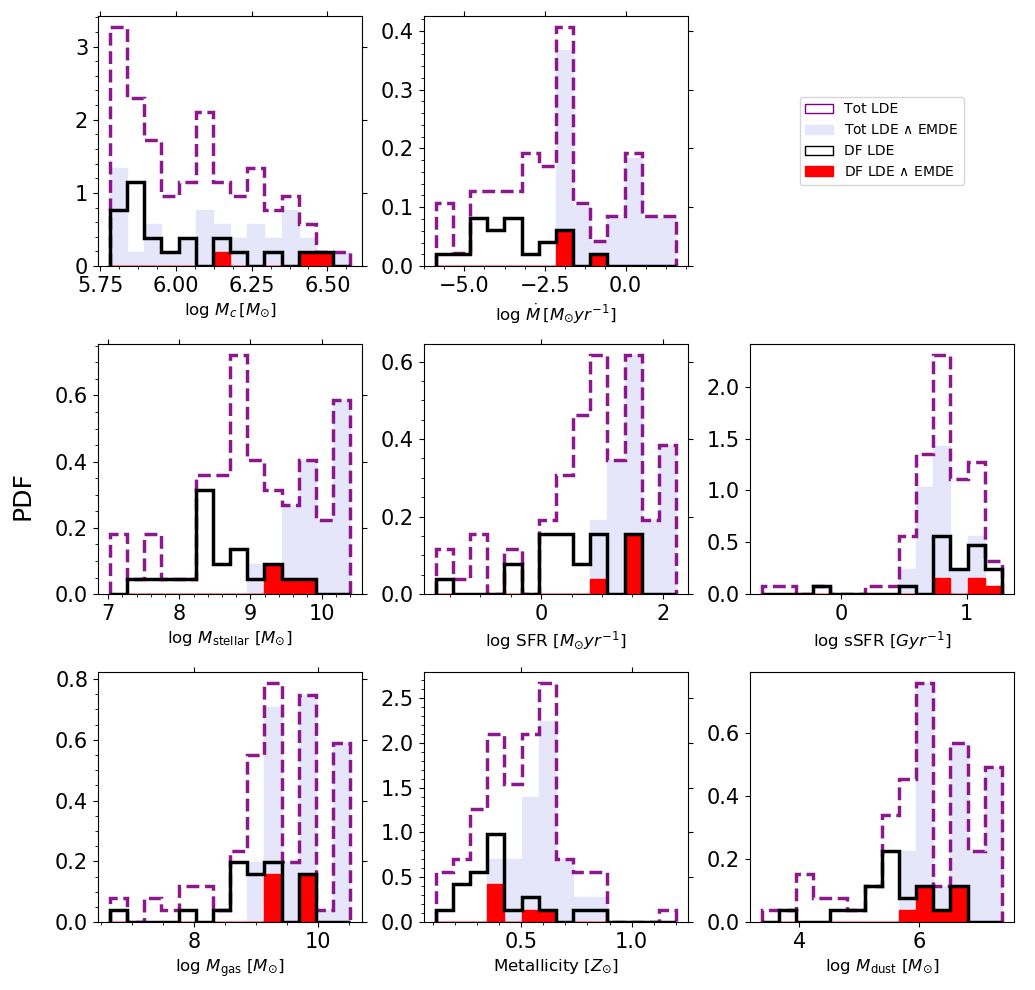}
         
    \caption{Probability Distribution Functions (PDFs) of different intrinsic properties of LDEs associated with a host galaxy before (purple dashed) and after (black solid) adding time delay due to DF at $z>6$. In each panel, the lavender bars represents the LDEs which are also EMDEs (see Sec.~\ref{sec:EMDEs}) for all events while the red region represents the same for the events which occur after considering time delays due to DF. \textit{Top left}: Chirp mass of LISA events \textit{Top middle}: total accretion rates of the LISA events. \textit{Middle left}:  stellar mass of the host galaxies. \textit{Middle center}: star formation rate of the host galaxies. \textit{Middle right}: specific star formation rate of the host galaxies. \textit{Bottom left}: gas mass of the host galaxies. \textit{Bottom middle}: distribution of the metallicities of the host galaxies.\textit{Bottom right}:  dust mass distribution in the host galaxies. Mass and metallicities are represented in solar units.}
    \label{fig:L_compint}
\end{figure*}
\section{Selection of LDEs for RT calculations} \label{selecRT}

In this section, among the GW events previously identified in \citetalias{Chakraborty_2023}, we select those that are both detectable with LISA (LISA Detectable Events, LDEs) and with EM follow-up observations (EM detectable Events, EMDEs). For what concerns LDEs, we compute the signal-to-noise ratio (SNR) of the GW emission from merging MBHs, and we select those events that are characterised by SNR$>5$  \citepalias[see Sec. 5.1 in ][]{Chakraborty_2023}. 
 To briefly summarize, in \citepalias{Chakraborty_2023}, we calculated the SNR of merging MBHBs over a given observational time $\tau$ as \citep{{Flanagan_1998}}:
\begin{equation}
\left (\frac{S}{N}\right)_{\Delta f}^2= \int_{f}^{f+\Delta f} d\ln f' \, \left[
\frac{h_c(f'_r)}{h_{\rm rms}(f')} \right]^2,
\label{eqSN}
\end{equation}
where $f_r$ is the GW rest-frame frequency, $f=f_r/(1+z)$ is the observed frequency, $\Delta f$ is the frequency
shift in the duration of $\tau$. Here, $h_c$ is the  characteristic strain which is the sky and polarization averaged strain amplitude over the number of cycles spent by MBHBs in the LISA bandwidth, and $h_{\rm rms}$ is the effective rms noise of the instrument. Recalling that LISA is sensitive at frequencies [10$^{-4}$--1.0]~Hz, we noted that a merger event can only be detected when the emitted GWs enter the LISA frequency range (even though the MBHB might still emit GWs outside this frequency window). The SNR corresponding to this minimum frequency limit is SNR$_{\rm thresh}$ which we considered to be 5 in order for a GW event to be detected by LISA.

\subsection{Intrinsic properties of systems hosting merging MBHs}\label{intrinsic_global}
Following the procedure described in Sec. \ref{G-BH}, we first associate to each GW event previously identified by \citetalias{Chakraborty_2023} the intrinsic properties of the galaxies hosting merging MBHs: the total (primary plus secondary) BH accretion rate (BHAR), the star formation rate (SFR),, the specific star formation rate (sSFR), the metallicity ($Z$), the dust mass ($M_{\rm dust}$), the stellar mass ($M_{\rm stellar}$) and the gas mass ($M_{\rm gas}$). We also show the chirp mass distribution of the merging MBHs in the same figure. In table \ref{intrinsicprops}, we report the intrinsic properties of the merger events to which a galaxy is associated without including dynamical friction (DF) effects. In what follows, we report the results with and without including DF effects. 

In Fig. \ref{fig:L_compint}, we show the Probability Distribution Function (PDF) of the intrinsic properties of all the LDEs assuming instantaneous merger (purple dashed line, Tot LDE). This PDF is normalized to one. For these events, we compute the time delay due to stellar DF (see Sec. 4.2 in \citetalias{Chakraborty_2023}). We then select only those MBHs whose merging is occurring at $z\gtrsim 6$ after having taken into account time delay effects. The black solid line denotes the LDEs among these events, called DF LDEs in the figure. In this case, for visual purposes, we normalize the PDFs to the fraction of LDEs that, after accounting for DF, are still occurring at $z\gtrsim 6$.

In the top left panel of Fig. \ref{fig:L_compint}, we show the chirp mass of LDEs before and after considering DF effect. We find that the average chirp mass of the total LDEs is $1.3 \times 10^6 M_{\odot}$ while adding the DF time delay, the average chirp mass decreases to $1.1 \times 10^6 M_{\odot}$.
From the top middle panel of Fig. \ref{fig:L_compint}, we can see that the total BHAR of LDEs in the \AGNfiducial{} run peaks at $\dot{M}\sim 0.01~\rm M_{\odot}~yr^{-1}$, and varies in the range $[\sim 10^{-6}-30~\rm  M_{\odot}~yr^{-1}]$. If considering the DF time delay effect, the highest BHAR of LDEs shifts to a lower value of around $\dot{M}\sim 0.1~\rm M_{\odot}~yr^{-1}$.  
The middle left panel depicts the stellar mass distribution of the host galaxies. We find an average of $M_{\rm stellar} \sim 4.8\times10^9~\msun$ ($\sim 8.5\times10^8 ~\msun$) without (with) including the delays due to DF.

The middle (right) panel of the second row, shows the distribution of the SFR (specific SFR, denoted as sSFR) in the host galaxies of LDEs. We find that the average SFR (sSFR) of LDEs in the \AGNfiducial{} is SFR$\sim 29~\msun~\rm yr^{-1}$ (sSFR$\sim 7.2~\rm Gyr^{-1}$), while it is $\sim 9\ \msun~\rm yr^{-1}$ ($\sim 9.3~\rm Gyr^{-1}$) if DF effects are taken into account. 
In the bottom left panel, we show the gas mass distribution of the host galaxies and see that the average $M_{\rm gas} \sim 6.7\times10^9~\msun$ ($\sim 2.2\times10^9 ~\msun$) without (with) including the delays due to DF.

 Finally, for what concerns the metallicities 
 ($Z$) and the dust masses ($M_{\rm dust}$), shown in the bottom center and bottom right panel respectively, we find average values of $Z_{\rm mean}\sim 0.5~Z_{\odot}$ ($\sim 0.4~Z_{\odot}$) and $M_{\rm dust} \sim 4.5\times10^6~\msun$ ($\sim 2\times10^6 ~\msun$) without (with) including DF effects. 
 
Overall, we find that the PDF of intrinsic properties computed according to the (i) instantaneous merging approximation are shifted towards higher values with respect to when (ii) DF effects are included. This occurs because in case (i), we are selecting events in the redshift range $6<z<9$ (see the left panel of Fig. \ref{fig:pdf_all}), while in case (ii) we lose all the events selected at $z\sim 6$ (the most evolved ones) and we are left only with the highest-$z$ events (see the right panel of Fig. \ref{fig:pdf_all}). The latters, being caught at earlier epochs in less evolved environments, are characterized by smaller values of BHAR, SFR, $Z$, $M_{\rm dust}$, $M_{\rm stellar}$ and  $M_{\rm gas}$.

 \begin{figure*}[hbt!]

        \includegraphics[width=2.0\columnwidth]{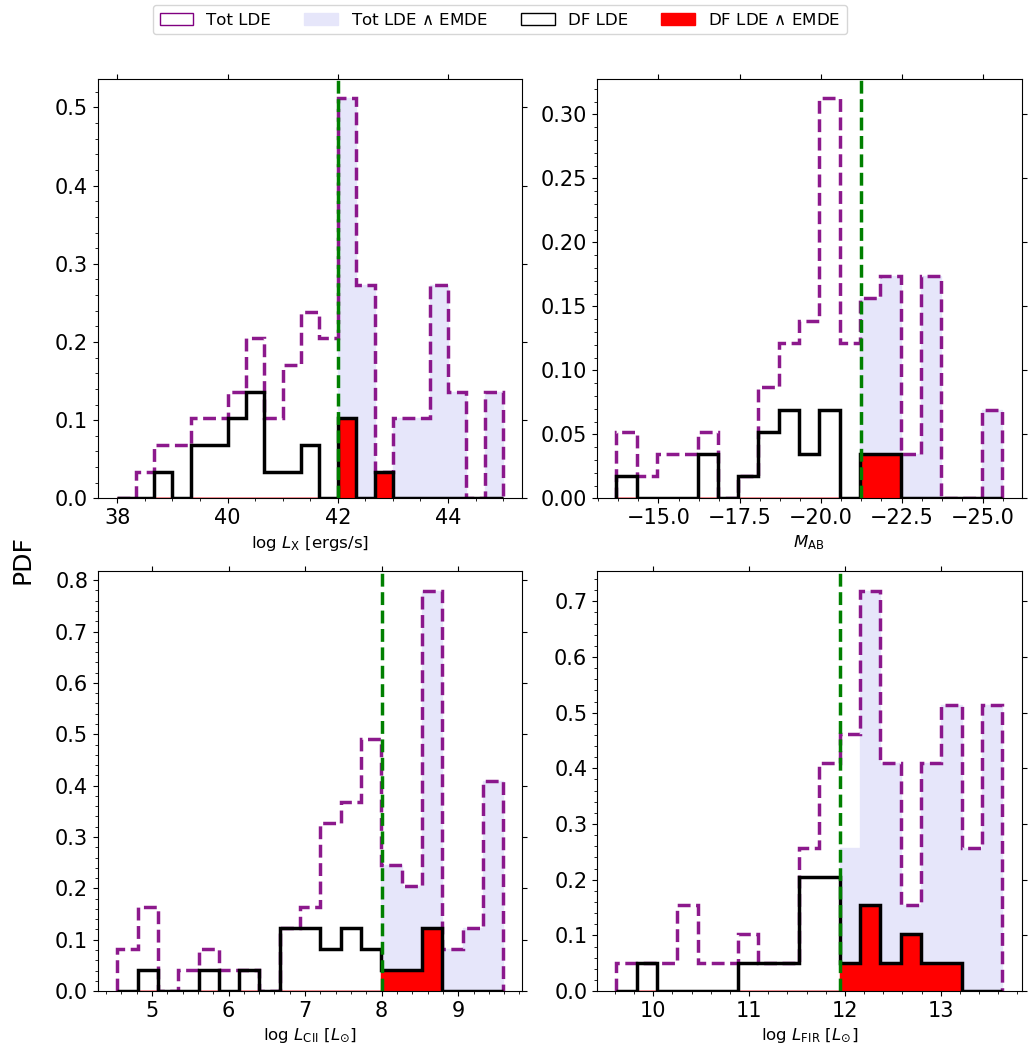}
       
    \caption{
    PDF comparison of different observable properties of LDEs (dashed purple), after adding time delay due to DF at $z>6$ (black solid) for follow-up observations in different wavelength bands: X-ray (upper left panel), UV (upper right), [CII] (bottom left), FIR (bottom right). The colours are the same as explained in Fig.\ref{fig:L_compint}. 
    }
    \label{fig:L_comp}
\end{figure*}

\begin{figure*}
        \captionsetup[subfigure]{labelformat=empty}
        \centering
        \subfloat[\centering ]{{\includegraphics[width=8.5cm]{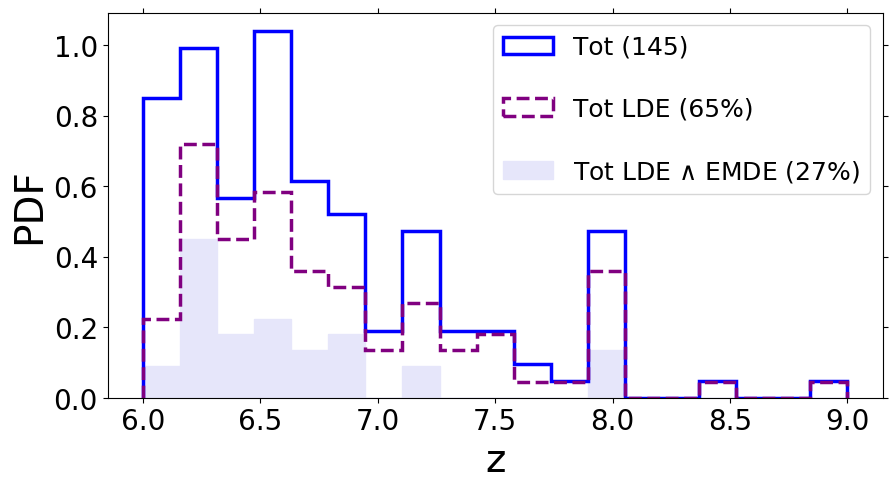} }}%
        \qquad
        \subfloat[\centering ]{{\includegraphics[width=8.5cm]{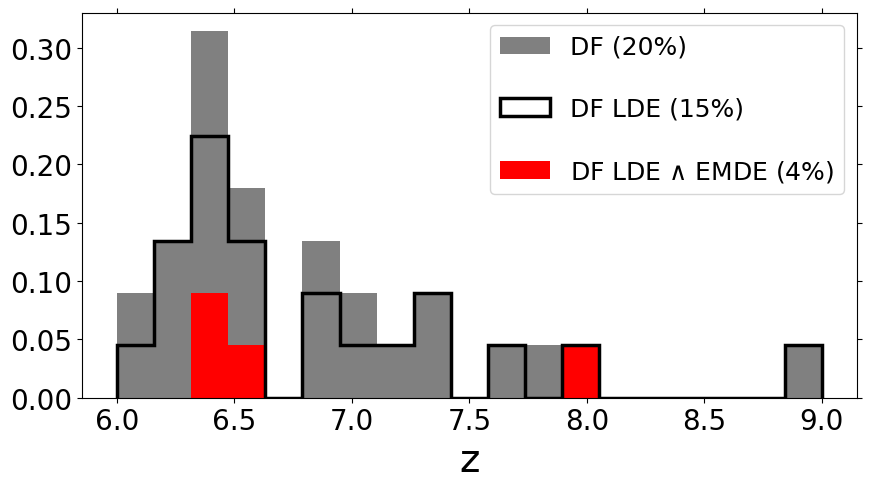} }}%

    \caption{ PDFs of all merging events and all LDEs (left panel) and DF events and DF LDEs (right panel). In the left panel, the blue solid line represents the total number of merging events in our \AGNfiducial{} run. The purple dashed line represents the LDEs in \AGNfiducial{} and the lavender area further denotes the LDEs which are also detectable in UV, X-ray, CII, and/or FIR bands by JWST, LynX, and ALMA. In the right panel, the grey area represents the mergers that take place at $z > 6$ after we consider time delay due to dynamical friction. The black solid line shows the LDEs of these aforementioned mergers while the red area represents the LDEs and EMDEs from these mergers.}
    \label{fig:pdf_all}
\end{figure*}
\begin{figure*}

        \includegraphics[width=2.0\columnwidth]{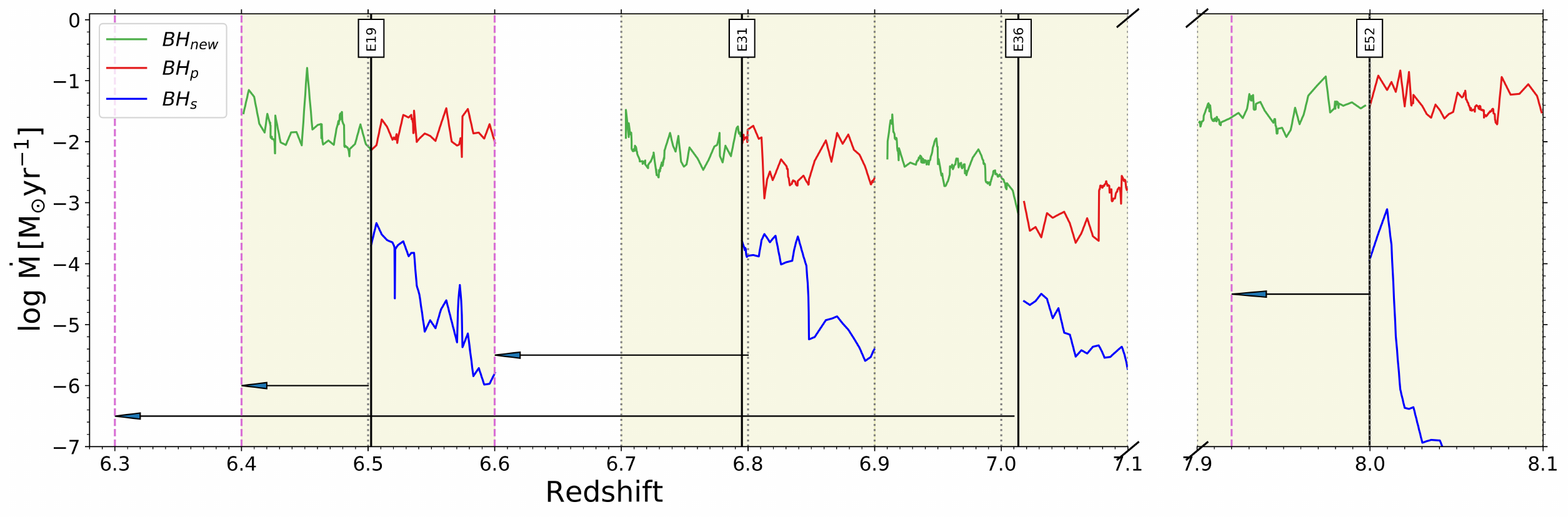}
        
    \caption{Redshift evolution of the MBH accretion rates ($\dot{M}$) for LDEs. We select 4 events for which the merger occurs at $z\gtrsim 6$ after adding time delays due to DF. For each event, the red (blue) line refers to the primary (secondary) MBH, labelled $BH_p$ ($BH_s$) in the figure, while the green line shows $\dot{M}$ of the new MBH resulting from the mergere ($BH_{new}$). The black solid vertical lines show the times at which the merger occurs in the simulation, according to the instantaneous merging approximation. The vertical pink-dashed lines show the new redshifts at which the BHs merge after adding delays due to DF. The black arrows show the change in the merger redshifts upon considering time delays due to DF in post-processing. The yellow patches show the time span for which the selected BHs evolve without undergoing any other mergers between two snapshots (the redshifts of the snapshots are depicted by vertical black-dotted lines).}
    
    \label{fig:mdot}
\end{figure*}

\subsection{Observable properties of systems hosting merging MBHs} \label{sec:EMDEs}

Starting from the intrinsic properties described above and following the formalism described in the Appendix \ref{MWL}, we compute the following observable properties: the X-ray luminosity ($L_{\rm X}$), the total (from stars and accreting BHs) UV luminosity ($L_{\rm UV}$) and the corresponding UV magnitude ($M_{\rm AB}$), the [CII] luminosity ($L_{\rm CII}$), and the far-infrared luminosity ($L_{\rm FIR}$). We report in table \ref{observable}, the observable properties of the galaxies to which a merger is associated (corresponding to the events reported in table \ref{intrinsicprops}). 

In Fig. \ref{fig:L_comp}, we show the PDFs of the aforementioned observational properties. The vertical lines represent the luminosity thresholds that we adopt to identify those sources that are more likely detectable with EM telescopes. In particular, for the UV, [CII], and FIR emissions, we consider the typical values found in the ALPINE survey \citep{lefevre2020} for $z\sim 5-6$ galaxies \citep{bethermin2020,Faisst2020,sommovigo2022}. For the X-ray, we consider a threshold of $L_X\sim 10^{42}$ erg~s$^{-1}$, that is both close to the confusion limit of Athena \citep{aird2013} and expected to be reached through a Lynx-like\footnote{https://wwwastro.msfc.nasa.gov/lynx/docs/LynxConceptStudy.pdf} instrument in $\sim 40$ ks \citep[see, e.g., Table 2 in][]{lops2023}.

Hereafter, we call "Electro-Magnetic Detectable Events" (EMDEs) those events whose observable properties are above at least one of the sensitivities of the EM telescopes considered. The red cyan-hatched region denotes those events that are both detectable with LISA and through EM telescopes, assuming instantaneous merger (Tot LDE $\wedge$ EMDE). The red cross-hatched region reports the EM AND LISA detectable events that are at $z\gtrsim 6$ after having considered DF effects. These are labelled in the figure as DF LDE $\wedge$ EMDE.

Considering DF effects, the fraction of LDE events that are EM detectable decreases from 31 \% to 21\%  in the Xray, from 26 \% to 18\%  in the rest frame UV, from 32\% to 21\% in [CII], from 45 \% to 32\%  in the FIR. In conclusion, only 20-30\% of the total LDEs in the \AGNfiducial{} run are also detectable with EM telescopes \footnote{In Appendix~\ref{nonmerg}, we have also compared the observable properties of host galaxies of LDEs with hosts of non-merging MBHs.}

\subsection{Final selection of LDEs}

\begin{figure*}[ht!]
	\centering
 \includegraphics[width=1.\textwidth] {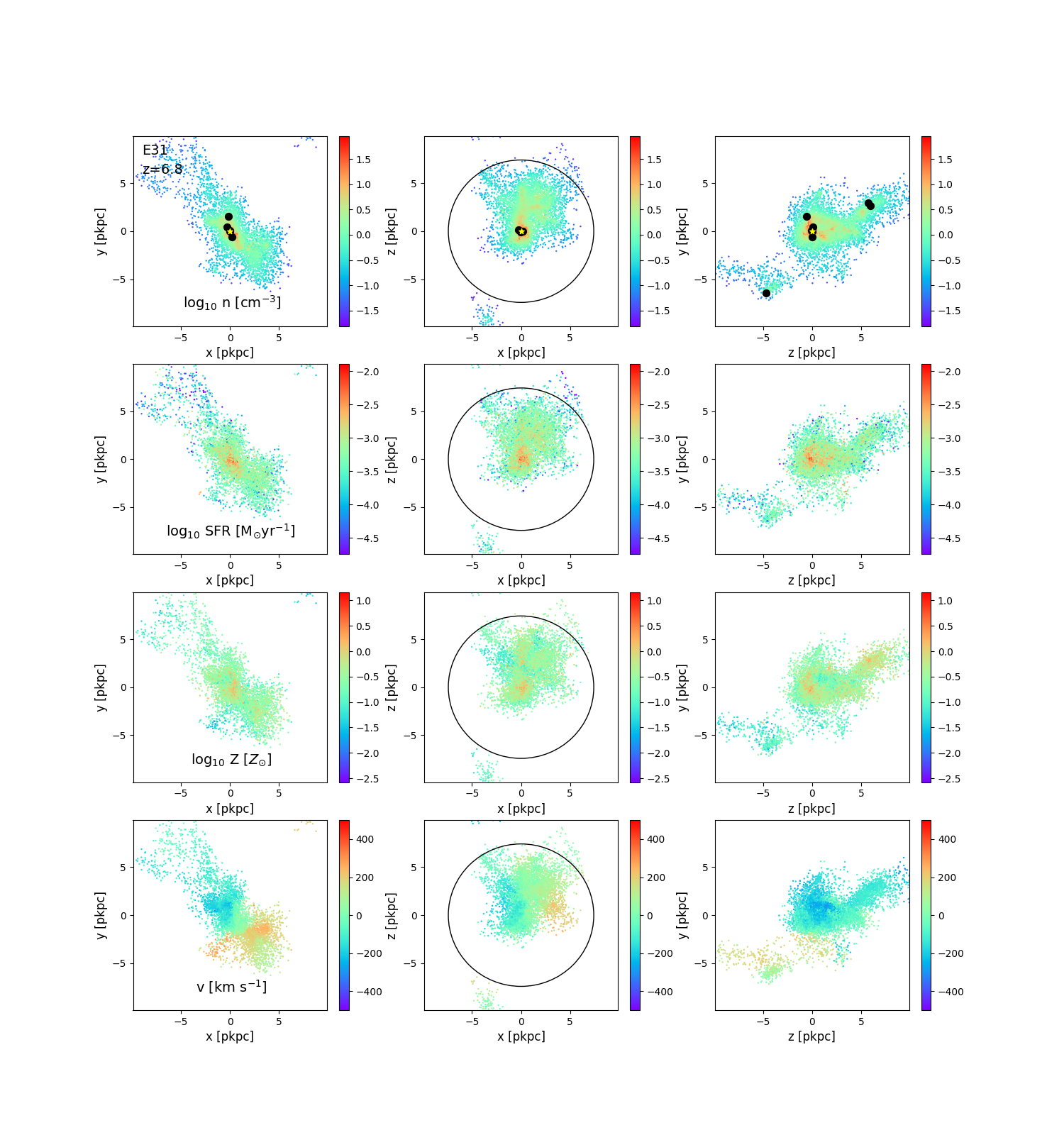}
    \caption{3D representation of the intrinsic properties for the event 31 (E31) at $z=6.8$. Each box has a side of $20$~pkpc and $1.2$~pkpc along the z/y/x directions (in the left/middle/right panel, respectively) for three different lines of sight (LOS). First row: gas density of the star-forming particles. The black circle identifies the region adopted to compute the intrinsic properties in Sec. \ref{intrinsic_global}, corresponding to 30\% of the virial radius. The filled black circles represent the location of the MBHs in the simulation; in particular, the location of the merger event is denoted by a yellow-filled star. The location of the MBHs and of the merger event remain the same for all rows. To avoid clutter and to provide a better view of the host galaxy intrinsic properties, filled black circles and the yellow-filled star are reported only in the first row. Second row: star formation rate of the SF particles. Third row: gas metallicity of the SF particles. Fourth row: gas velocity of the SF particles along the different LOS considered.}
    \label{fig:E31_maps_int}
\end{figure*}

The redshift distribution of the events as selected based on their observability with LISA and EM telescopes is shown in Fig. \ref{fig:pdf_all}. In the left panel, all events (Tot) are shown with a blue-shaded region, the LDEs (Tot LDEs) with a cyan-shaded region, and LDEs detectable with EM telescopes (Tot LDE $\wedge$ EMDE) with a red cyan-hatched region. In the right panel, the PDFs of those MBHs whose merging is occurring at $z\gtrsim 6$ after including time delay effects (DF) are shown with a dark-violet shaded region, the LDEs among the DF events (DF LDEs) with a lilac shaded region, the EM detectable events among the DF LDEs with a red cross-hatched region (DF LDE $\wedge$ EMDE). 

We end up with a total of 4 events that are interesting from the point of view of their observational properties, being the brightest in different electromagnetic bands: ($L_{X}$, $\rm L_{UV}$, $\rm L_{CII}$, and $\rm L_{FIR}$)
E19 at $z=6.5$ , E31 at $z=6.8$ , E36 at $z=7.0$  and E52 at $z=8$. In particular, we find E19 to be the brightest in $\rm L_{CII}$ and $\rm L_{FIR}$ bands and E52 in $\rm L_{UV}$ and $L_{X}$ band. 

Fig. \ref{fig:mdot} shows the accretion rate evolution of the MBHs involved in these events. We remind that these events are characterised by the following properties: (i) they are detectable with LISA with a SNR$>$5; (ii) they are detectable with EM telescopes in at least one band (see Fig. \ref{fig:L_comp}); (iii) the merging of the MBHs is expected to occur at $z\gtrsim 6$ after considering the time delay due to dynamical friction.

In Fig. \ref{fig:mdot} we show the accretion rate of the primary and secondary MBH before the merger, and the accretion rate of the MBH after the merger, namely after that the primary MBH has swallowed the secondary one. The final MBH has thus inherited the location of the primary BH, while its total mass is given by the sum of the primary plus secondary. We see that the accretion rate after the merger event mostly follows the accretion rate of the primary MBH before the merger; its larger mass as compared to the secondary MBHs is thus always dominating the accretion rate of the system.

\subsection{3D representation of intrinsic properties}\label{3Dintr}

We finally showcase the spatial distribution of the intrinsic properties of the selected events at the snapshot closest to the numerical merger time\footnote{Ideally, for having a proper description of the host galaxy and MBH properties of a merger event, we should consider the merger time after having included DF effects. However, given the assumption of instantaneous merger, the last time the galaxy and the MBH properties were self-consistently computed is at the numerical merger time. For this reason, it is more physical to consider the snapshot closest to the numerical merger time.} Fig. \ref{fig:E31_maps_int} shows the gas density (top panels), the SFR (second row), gas metallicity (third row), and the gas velocity (bottom panels) of the star-forming particles relative to the event 31 along three different lines of sight (LOS).

In Fig. \ref{fig:E19_maps_int}, \ref{fig:E36_maps_int}, and \ref{fig:E52_maps_int} we report the same properties for the other three selected events. The filled black circles represent the location of the BHs in the simulation; in particular, the location of the merger is denoted by a yellow-filled star. The solid circle identifies the region adopted to compute the intrinsic properties in Sec. \ref{intrinsic_global}, namely 30\% of the virial radius, hereafter called "SF region": $r_{\rm SF}=8.2, 7.4, 5.0, 4.8$~kpc, for E19, E31, E36, E52, respectively).

These figures clearly show that the location of the star-forming particles coincides with regions of relatively high density (log$_{10} n\gtrsim 1-2~ \rm cm^{-3}$), which thus become the most metal-enriched ones  (log$_{10} Z\sim 0.1-1~ \rm Z_{\sun}$), reaching the solar metallicity in the densest regions. In particular, for what concerns E31, the primary BH is located at a distance smaller than/close to the smoothing length ($\sim 24-59$ pc) from the densest (log$_{10} n=2.7~\rm cm^{-3}$), more star forming (log$_{10} SFR=-1.5~\rm M_{\odot}~yr^{-1}$), and metal-enriched (log$_{10} Z\rm[Z_{\odot}]=0.58$) particle.

From the bottom row, we notice that along the z and y directions, the gas velocities of the star-forming particles resemble the dynamics of a rotating disk in the edge-on view. The dynamics of the star-forming particles will be further analysed in Sec. \ref{halpha}. Here, we underline that a disk-like velocity distribution is also visible in E52 along the x direction and that these findings are in agreement with recent ALMA results \citep{Rowland2024}.

\section{EM signals from selected events}\label{RT}

\begin{figure}
	
        \centering
       
        \includegraphics[width=0.79\columnwidth]{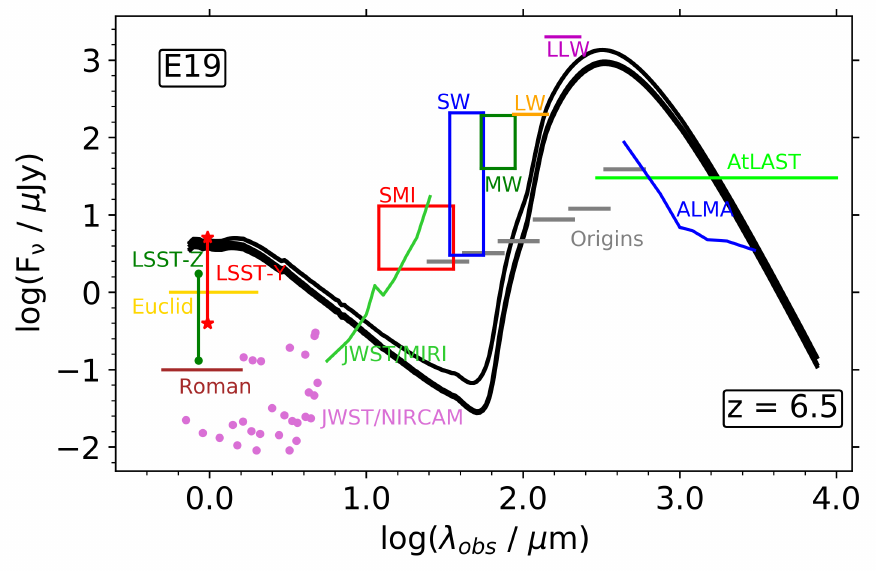}\\
        \includegraphics[width=0.79\columnwidth]{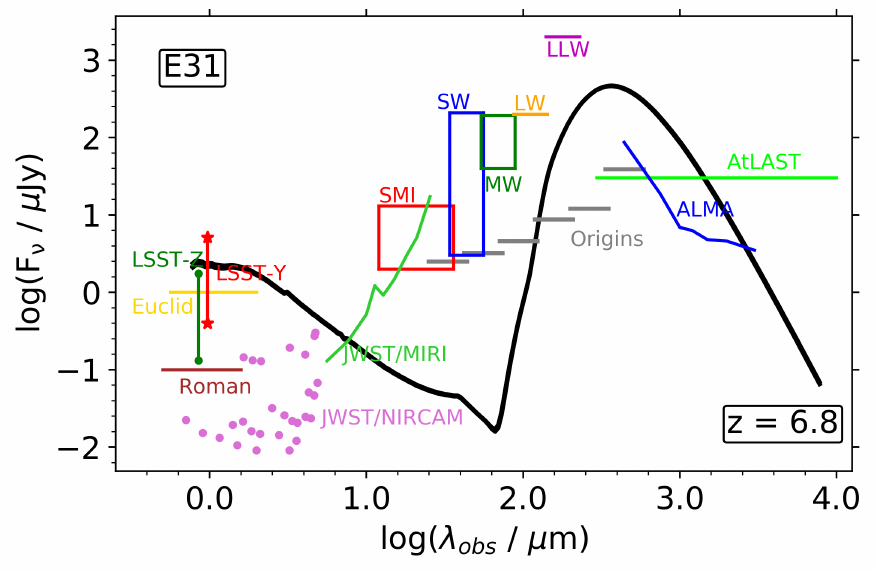}\\
        \includegraphics[width=0.79\columnwidth]{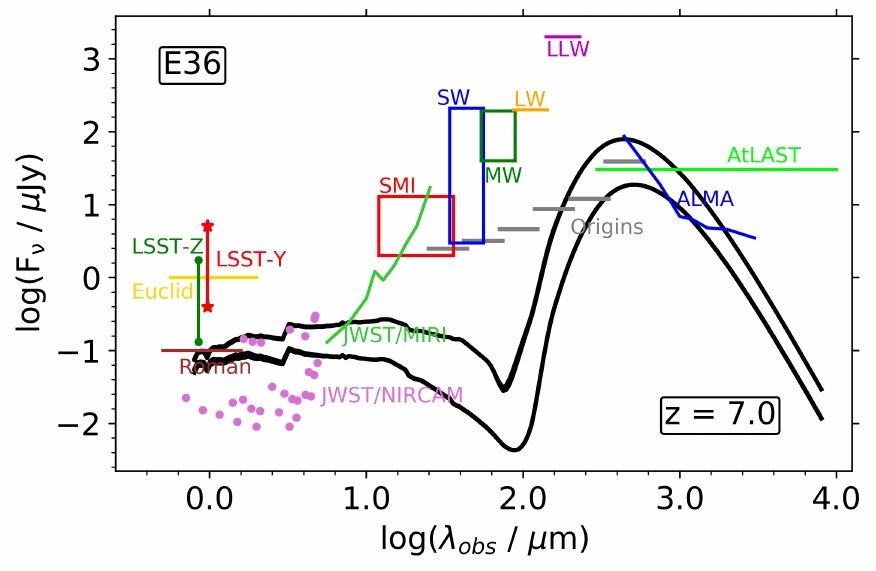}\\
        \includegraphics[width=0.79\columnwidth]{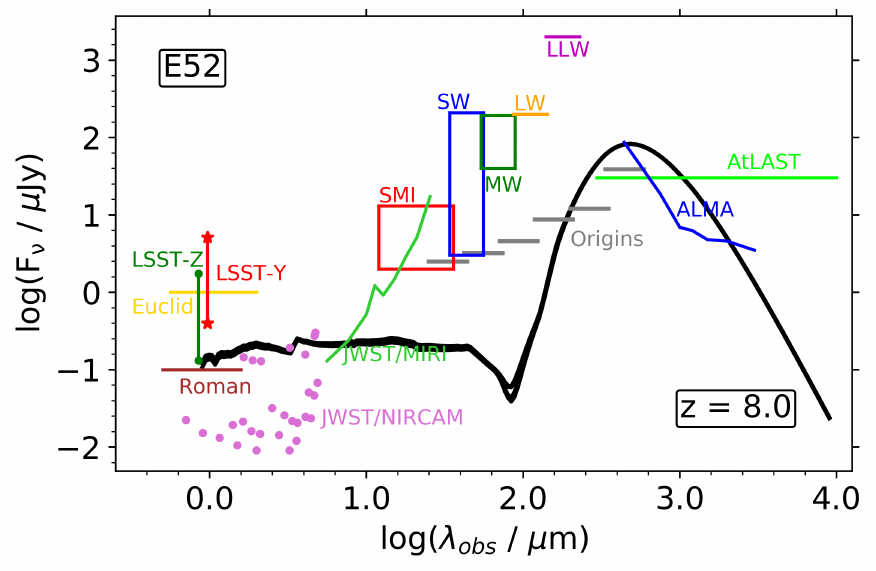}
       
    \caption{Spectral energy distributions (SEDs) of the selected events: E19 at $z=6.5$ (first row), E31 at $z=6.8$ (second), E36 at $z=7.0$ (third), E52 (last) at $z=8.0$, extracted from different lines of sight shown as black solid lines in Fig. \ref{fig:E31_maps_int}, from an aperture containing the SF region ($r_{\rm SF}=8.2, 7.4, 5.0, 4.8$~kpc, respectively). The synthetic SEDs are compared with the sensitivities of the different EM telescopes: LSST filters Z (dark green vertical line) and Y (red vertical line), Roman (brown horizontal line), JWST/NIRCAM (orchid dots), JWST/MIRI (green curved line, for an exposure time of $\sim$ 3 hrs), Origins-like (gray horizontal lines, at 5$\sigma$ in 1 hour), SPICA-like (red, blue, green boxes, yellow, magenta horizontal lines at 5$\sigma$ in 1 hr represented by the top portions of the rectangles and at 3$\sigma$ by the lower sides of the rectangle; if the confusion limit is achieved within an hour, then the sensitivies are depicted by the lines), ALMA (blue curved line, 10 hrs) and AtLAST (light green horizontal line) are shown.}
    \label{fig:6bsed}
\end{figure}

We perform RT calculations following the set-up described in Sec. \ref{RTsim} and considering the simulated volume centered on the galaxies hosting the events selected in Sec. \ref{selecRT}.

\subsection{Synthetic spectra}
In Fig. \ref{fig:6bsed}, we show the SEDs resulting from our RT calculations for the selected events, and we compare them with the sensitivities of different EM telescopes. 
From this comparison, it results that E19 and E31 are detectable at short ($\lambda_{\rm obs} \lesssim 10~\rm \mu m$) and long ($\lambda_{\rm obs} \gtrsim 100~\rm \mu m$) wavelengths with any of the current/planned EM telescopes sensitive to these wavelengths (e.g. LSST, Euclid, Roman, JWST, and Origins, ALMA, AtLAST, respectively), whereas they are far below the detection limit of a SPICA-like telescope (see also PRIMA\footnote{https://prima.ipac.caltech.edu/}). However, we underline that our model does not include any dusty torus emission. The detectability of LDEs at these wavelengths thus depends on the possible presence of the torus and on its properties \citep[i.e. dust mass and temperature, see e.g. Fig. 11 in][]{DiMascia:2021}. Similar considerations apply to E36 and E52, but these events will be detectable only by JWST (and barely by Roman).

\subsection{Synthetic maps of E31}\label{maps}
\begin{figure*}

        \centering
        \hspace*{1cm}
        
        \includegraphics[width=0.8\textwidth, height=0.4\textheight]{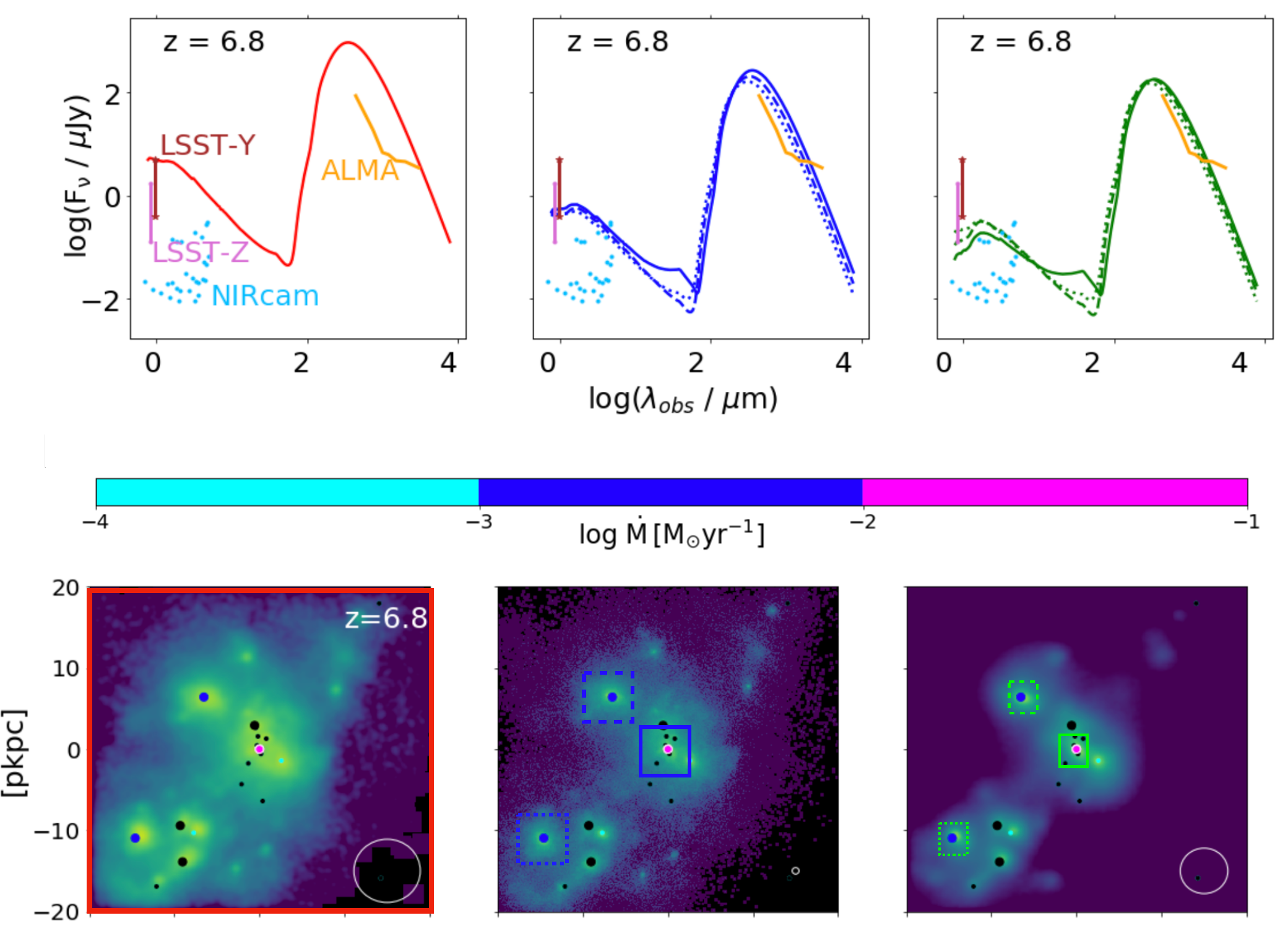}
        \includegraphics[width=0.8\textwidth, height=0.4\textheight]{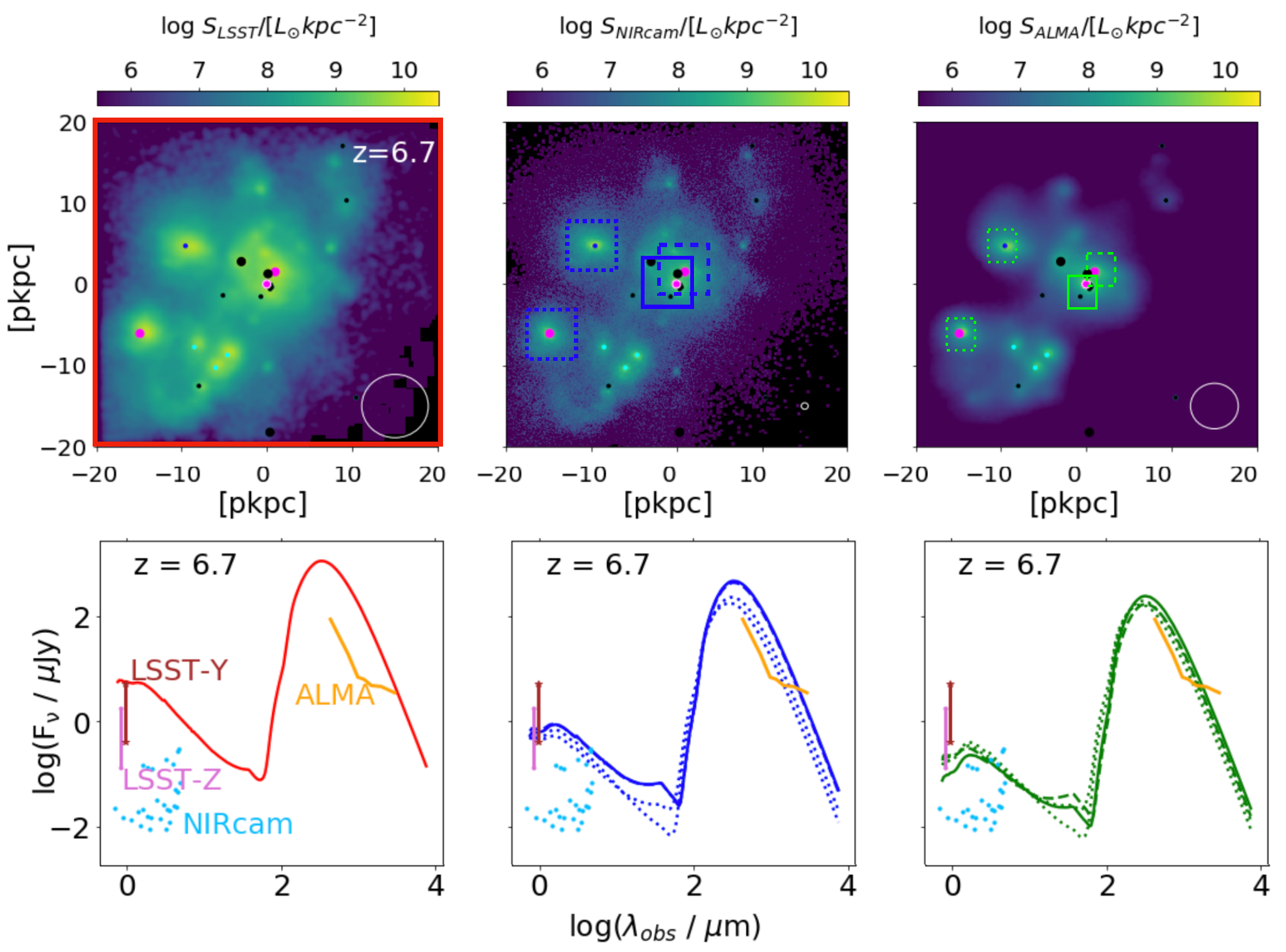}

    \caption{Simulated maps (second and third row) in different observable bands and SEDs (first and bottom row) relative to event 31. From left to right we show the maps in the wavelength range of LSST (0.9-1.1 $\mu$m), NIRcam (3.8-5.1 $\mu$m), and ALMA (0.8-1.1 mm). Each map is convolved to match the angular resolution of the corresponding instruments considered: LSST (0.7''), NIRCAM (0.07''), and ALMA (0.5''). These values correspond to beam sizes of approximately 4 kpc, 0.4 kpc, and 3 kpc, respectively, at $z \sim$ 6.8, and are shown as white circles at the bottom of each panel in the second and third rows. 
    The circles with white edgecolor represent the coalescing MBHs while the others represent the non-coalescing MBHs in the same halo. MBHs are coloured based on their mass accretion rates: MBHs with $\dot{M} <10^{-4} ~\msun$ are represented in black, $10^{-4} ~\msun \leq \dot{M} <10^{-3} ~\msun$ in cyan, $10^{-3} ~\msun \leq \dot{M} <10^{-2} ~\msun$ in blue, and MBHs with $\dot{M} >10^{-1} \msun$ in magenta. Small, medium and large circles represent MBHs with masses $\rm M_{BH} \leq 10^5 ~\msun$, $10^{5.5}~\msun \leq \rm M_{BH} \leq 10^{6}~\msun$ and $\rm M_{BH} > 10^6 ~\msun$, respectively.
    The top and bottom rows show the SEDs at $z$ = 6.8 and $z$ = 6.7, respectively. The left panel indicates the emission relative to the entire field of view. In the middle and right panels, a square region with size 2 kpc and 800 pc respectively is selected around the merging BH (solid contour in the emission maps) to extract the SED (solid line in the SED). The same is done for other AGN in the field of view (dashed contour in the emission maps and dashed line in the SED). 
    }
    \label{fig:maps31_con}
\end{figure*}
Once we have checked that the selected events are detectable in one or more EM bands, we investigate whether their emission properties differ from the ones of a typical AGN, powered by a single MBH. For this kind of study, we focus our attention on a specific event, e.g. E31. This event is predicted to occur at $z=6.8$ ($z=6.6$) assuming instantaneous merger (considering the time delay due to stellar dynamical friction), as can be deduced from Fig. \ref{fig:mdot} (see the solid black vertical line associated to E31, and the dashed magenta vertical line indicated by the arrow).

In the panels of Fig. \ref{fig:maps31_con}, the second and third rows show maps of the continuum emission predicted by our RT calculations for E31 in the wavelength ranges covered by LSST, JWST, and ALMA (left-most, middle, and right-most panels, respectively). In the same panels, coloured (black) circles represent the position of MBHs accreting with $\dot{M}$ larger (smaller) than $10^{-4} ~\msun~\rm yr^{-1}$. MBHs involved in the merger event are highlighted with white circles.  
 
We also show in the top and bottom panels the SEDs relative at $z$ = 6.8 and $z$ = 6.7, namely at redshifts before and after the merging. In particular, we show the emission from the entire simulated field of view (left panel), the emission associated with the merging BH and with other representative individual AGNs (two at $z=6.8$ and three at $z=6.7$) in the field of view (middle and right panels). 
By comparing the SEDs of the post-merger BH against the ones of isolated AGN we find only marginal differences in the synthetic spectra, mostly in the mid infra-red band. However, in this region, the flux is below the detection limit of current/planned telescopes sensitive to this wavelength range.

We conclude that it will be impossible to identify a LDE from the continuum SEDs because of the absence of specific signatures from the merging MBHs.

\begin{figure*}    
\includegraphics[width=0.3\textwidth,height=\textheight,keepaspectratio]{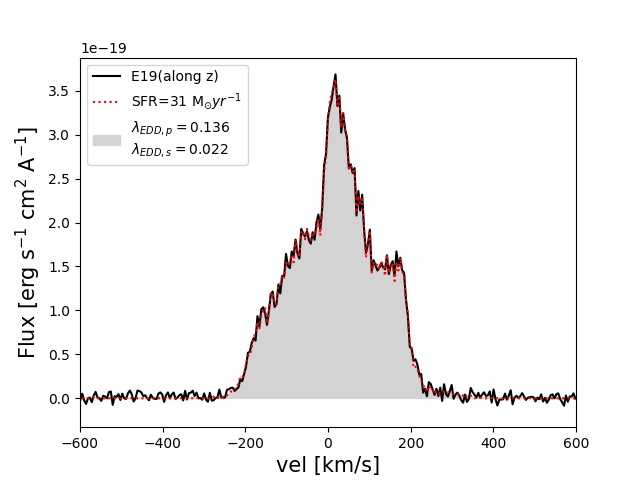}
\includegraphics[width=0.3\textwidth,height=\textheight,keepaspectratio]{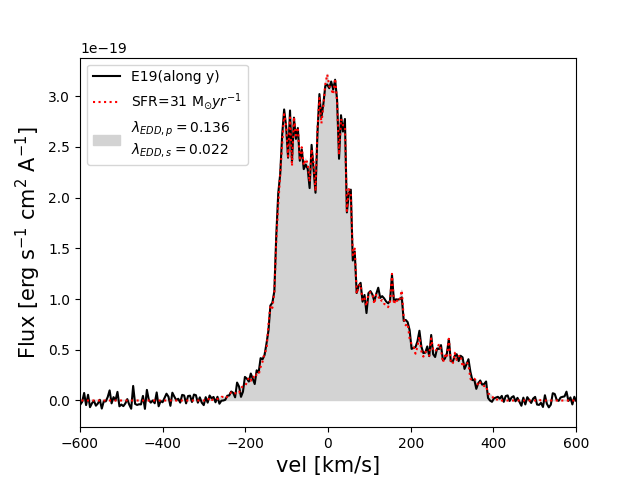}
\includegraphics[width=0.3\textwidth,height=\textheight,keepaspectratio]{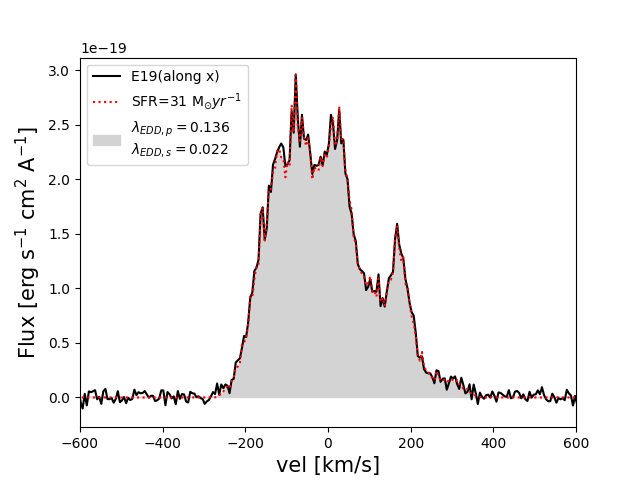}\\
\includegraphics[width=0.3\textwidth,height=\textheight,keepaspectratio]{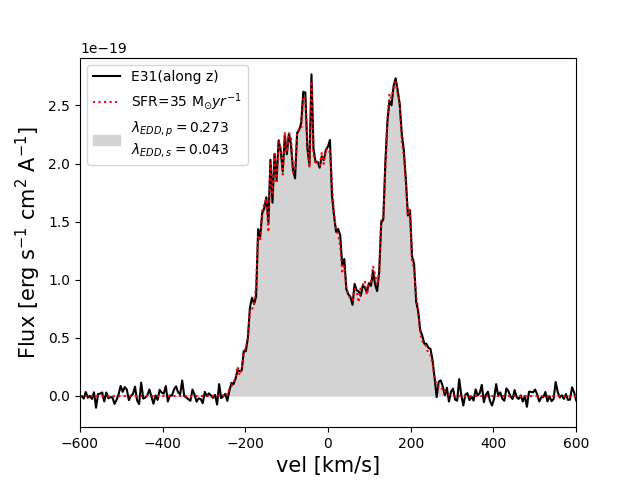}
\includegraphics[width=0.3\textwidth,height=\textheight,keepaspectratio]{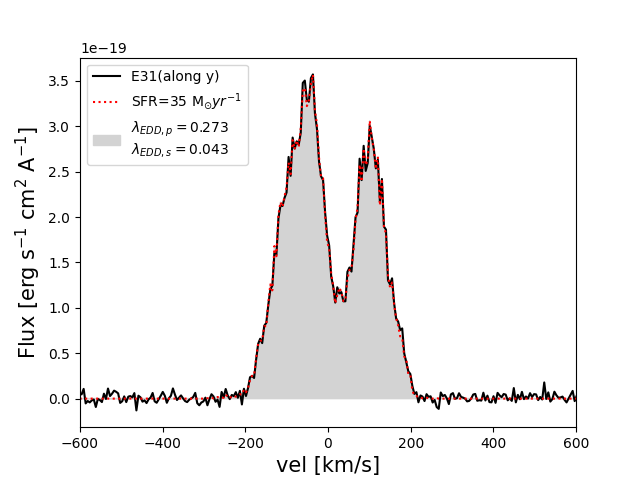}
\includegraphics[width=0.3\textwidth,height=\textheight,keepaspectratio]{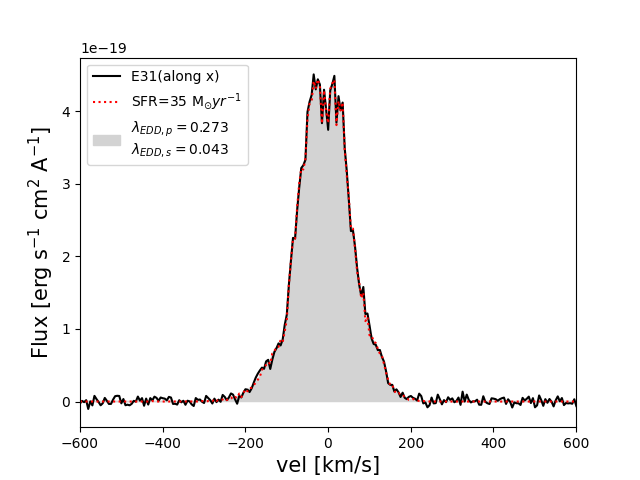}\\
\includegraphics[width=0.3\textwidth,height=\textheight,keepaspectratio]{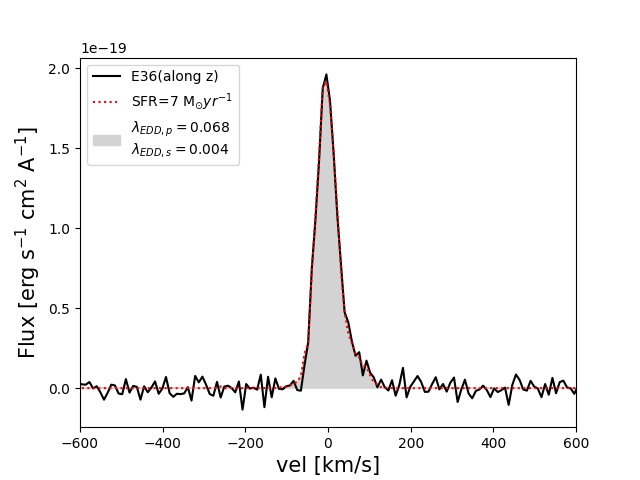}
\includegraphics[width=0.3\textwidth,height=\textheight,keepaspectratio]{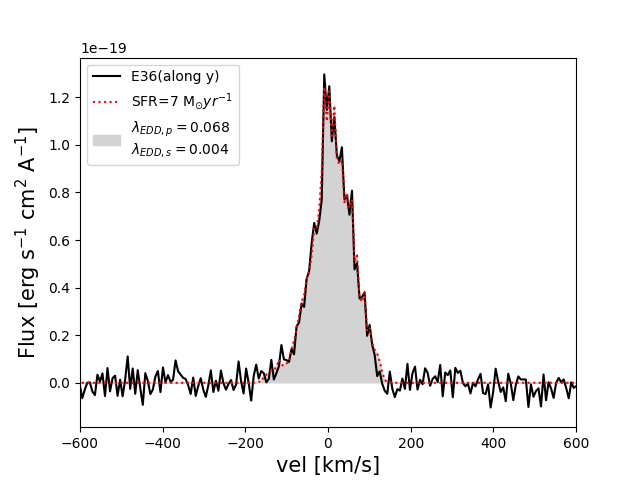}
\includegraphics[width=0.3\textwidth,height=\textheight,keepaspectratio]{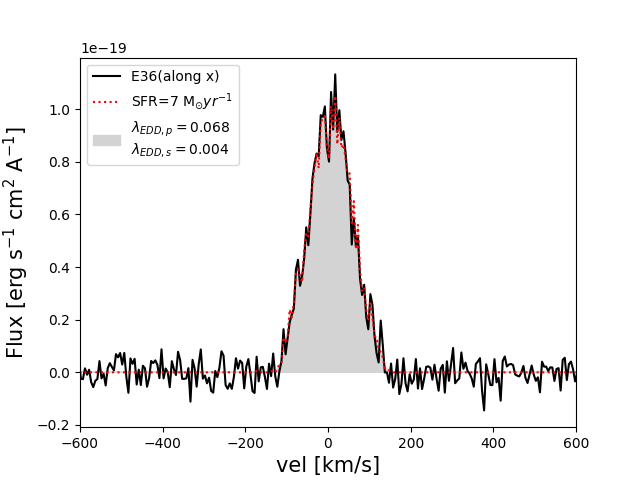}\\
\includegraphics[width=0.3\textwidth,height=\textheight,keepaspectratio]{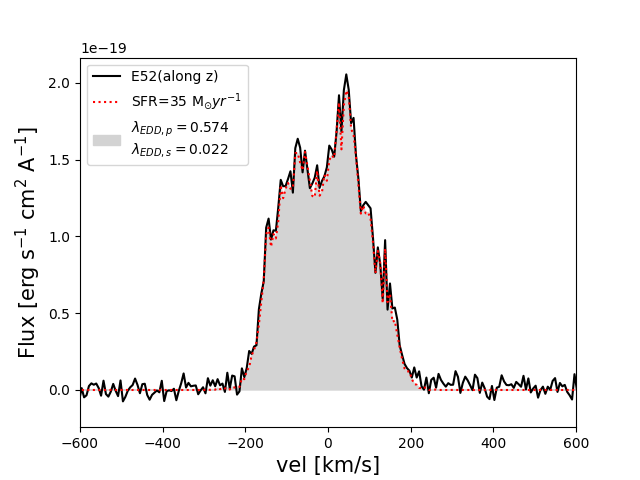}
\includegraphics[width=0.3\textwidth,height=\textheight,keepaspectratio]{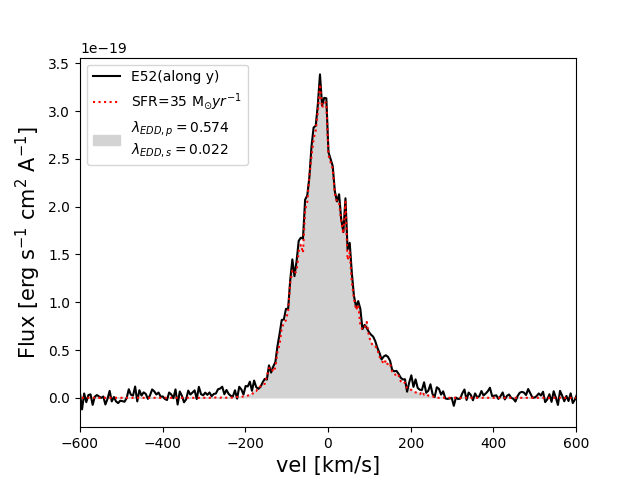}
\includegraphics[width=0.3\textwidth,height=\textheight,keepaspectratio]{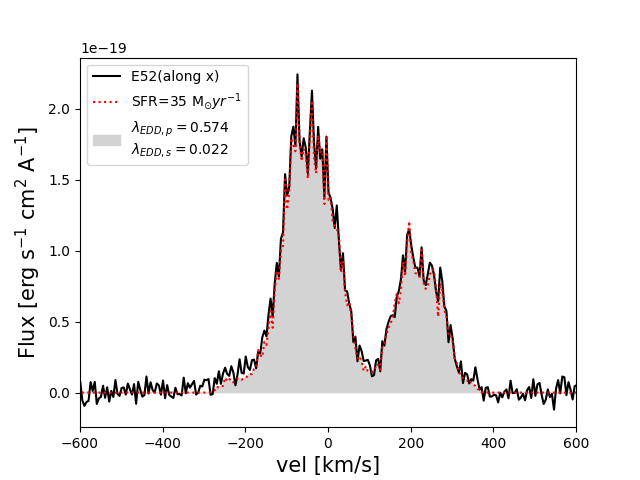}\\
    \caption{Predicted H$_{\alpha}$ line profiles from simulations for E19, E31, E36 and E52 (first, second, third, and last rows, respectively) shown along LOS in the z, y, and x (first, second and last columns, respectively) directions. The grey-shaded regions represent the calculated results. The solid black lines indicate the total flux (from SF, the accreting primary and secondary MBHs) along with the noise component and the red-dotted lines show the contribution only from SF. }
    \label{fig:Lalpha_E31}
\end{figure*}

\section{\texorpdfstring{H$_{\rm \alpha}$} ~~emission line from merging MBHs}\label{halpha}

\begin{figure*}
\centering
\includegraphics[width=0.42\textwidth]{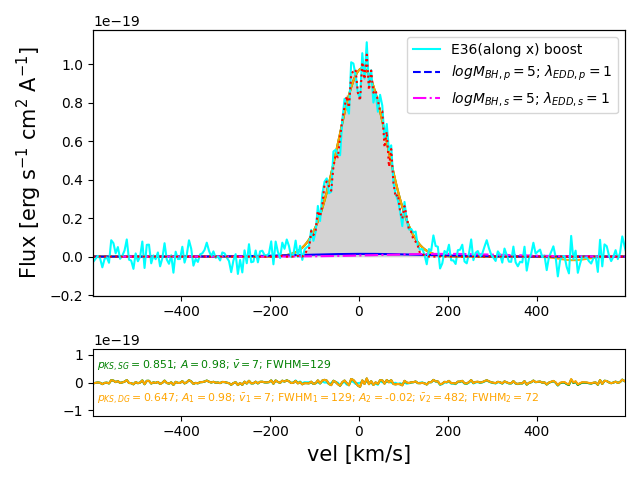}
\includegraphics[width=0.42\textwidth]{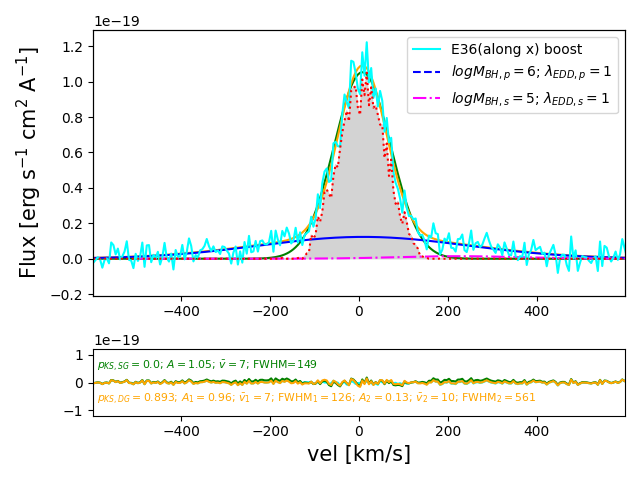}\\
\includegraphics[width=0.42\textwidth]{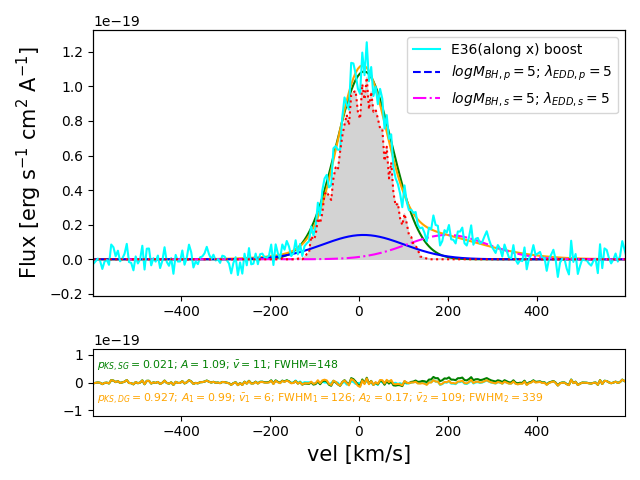}
\includegraphics[width=0.42\textwidth]{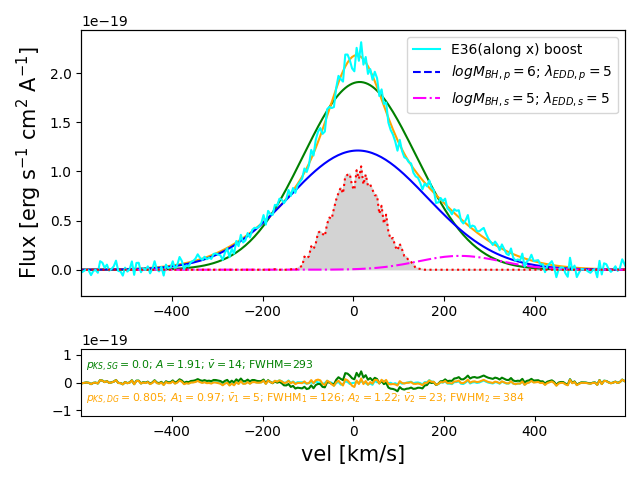}\\
\includegraphics[width=0.42\textwidth]{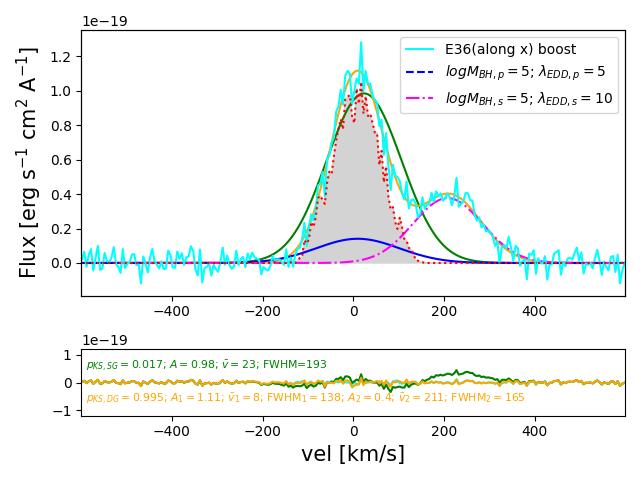}
\includegraphics[width=0.42\textwidth]{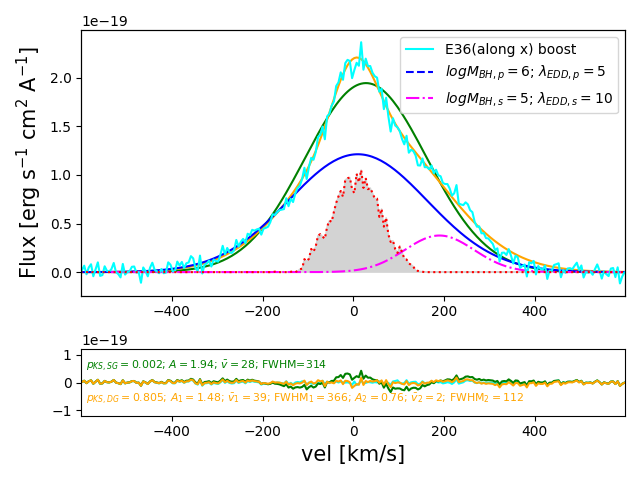}\\
\includegraphics[width=0.42\textwidth]{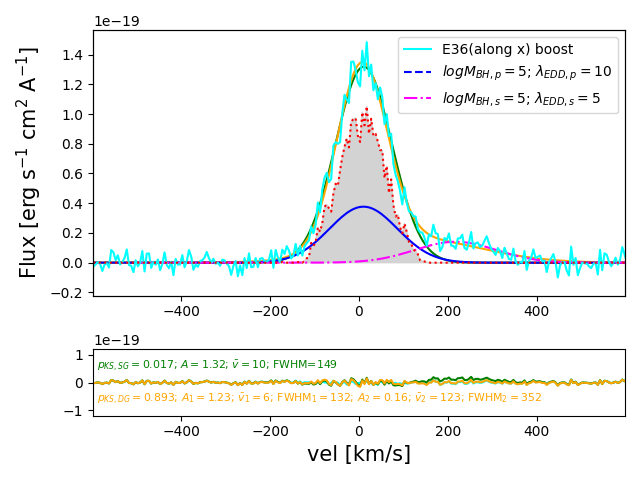}
\includegraphics[width=0.42\textwidth]{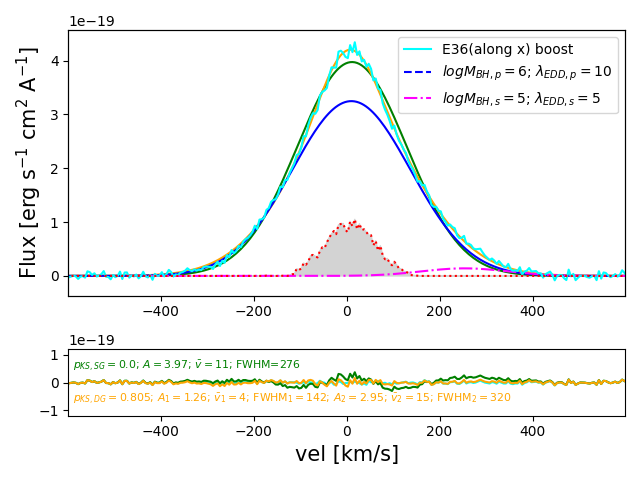}\\
    \caption{Re-simulated H$_{\rm \alpha}$ profile for E36 in the LOS along the x direction. The left column shows results for constant masses of MBHs but varying accretion rates. For the same values of the accretion rates as shown in the left column, we show the results when the primary MBH is ten times more massive than the secondary MBH. The blue (pink) dashed line represents the contribution from primary (secondary) MBH. The bottom panel in each figure refers to the added random noise in the synthetic spectrum along with the residual of the best-fitting model as explained in Sec.~\ref{boost}.  }
    \label{fig:Lalpha_E36_boost}
\end{figure*}

High-$z$ dual AGN recently suggested by JWST observations are discovered by the means of double-peaked, broad emission Balmer lines \citep[e.g.][]{maiolino2023b,ubler2023}. To investigate whether it is possible to infer the presence of the accreting and merging MBHs in our selected events, we compute their H$_{\rm \alpha}$ emission lines. The H$_{\rm \alpha}$ line arises from recombinations occurring in the ISM of the host galaxy and in the broad line region (BLR) of the accreting BHs, producing a narrow and two broad components, respectively.  

The luminosity of the narrow component (in units of erg~s$^{-1}$) scales with the galaxy SFR (in units of $\rm M_{\odot}~yr^{-1}$), following \citet{kennicutt2012}:
\begin{equation}\label{SFR_alpha}
L_{\rm H_{\rm \alpha}}^{\rm SFR}=1.86\times 10^{41} \rm SFR.
\end{equation}
For what concerns its profile, it is shaped by the velocity distribution of the SF particles. To compute the profile of the line resulting from the star formation, we first calculate the PDF of the velocities of the SF particles, weighted by their SFR, and then we re-normalise it in such a way that its integrated flux corresponds to $L_{\rm H_{\rm \alpha}}$, as per Eq. \ref{SFR_alpha}.

The luminosity of the broad components ($L_{\rm H_{\rm \alpha}}^{\rm M_{\rm BH}}$) is instead proportional to the luminosity of the accreting BHs at 5100~$\angstrom$~($L_{\rm5100}$), as in eq. 2 by \citet{Reines2013}:
\begin{equation}
L_{\rm H_{\rm \alpha}}^{\rm M_{\rm BH}}=5.25\times 10^{42}\left (\frac{L_{\rm5100}}{10^{44}~\rm erg~s^{-1}}\right)^{1.157}~\rm erg~s^{-1},
\end{equation}
where $L_{\rm5100}=f_{\rm bol} L_{\rm bol}\sim f_{\rm bol}\epsilon \lambda_{\rm EDD}\dot{M}_{\rm EDD}c^2$, $f_{\rm bol}=0.1$ \citep{lusso2012}, and $\lambda_{\rm EDD}$ is the Eddington ratio, namely the ratio between $\dot{M}$ and the Eddington accretion rate $\dot{M}_{\rm EDD}$ of each BH. The full width at half maximum of the broad components (FWHM$_{\rm BH}$) is finally related to the above quantities through eq. 5 by \citet{Reines2013}:
\begin{equation}
\log \frac{M_{\rm BH}}{\rm M_{\odot}} =  6.6 +0.47 \log \left(\frac{L_{H_{\rm \alpha}}^{\rm M_{\rm BH}}}{10^{42}~\rm erg~s^{-1}}\right)+2.06~\log \left(\frac{FWHM_{\rm BH}}{10^3~\rm km~s^{-1}}\right).
\end{equation}

The line centroids associated with the BLR of the MBHs are shifted with respect to each other based on their relative velocity. In particular, their relative position in the velocity space depends on the following relative velocities: we call $v_{\rm gal}-v_p$ the relative velocity between the galaxy and the primary BH, and $v_p-v_s$ the relative velocity between the merging BHs.

To summarise, the profile of the H$\alpha$ line is determined by the following quantities: SFR, velocity distribution of the SF particles, $\lambda_{\rm EDD,p}$, $M_p$, $\lambda_{\rm EDD,s}$, $M_s$, $v_{\rm gal}-v_p$, and $v_p-v_s$. All these properties are predicted by our simulation\footnote{The relative velocities between the primary and secondary MBHs were computed at those timesteps when the first condition for the merging is satisfied but not the second one (see condition (i) and (ii) in Sec. \ref{BH-model}, respectively).} that allows us to properly compute the H$_{\rm \alpha}$ emission line arising from our selected events.

\subsection{Synthetic H$_{\rm \alpha}$ emission line from the selected events}

We report in Fig. \ref{fig:Lalpha_E31} the H$_{\rm \alpha}$ line predicted by our simulations for the 4 selected events (E19, E31, E36, E52, from the top to the bottom), along different LOSs (along the z, y, and x direction in the left, middle, and right panel, respectively). The shaded region represents the result of our calculations, the solid black line denotes the total flux from the three components (the SF, and the two accreting BHs) plus noise\footnote{We add to our simulated spectra, on velocity bins of $\Delta v=5~\rm km~s^{-1}$, a random number extracted by a Gaussian distribution having mean equal to zero and standard
deviation equal to $5\times 10^{-21} \rm erg~s^{-1}~cm^{-2}~A^{-1}$.}, the red dotted line shows the contribution from the SF. 

We notice that, for what concerns E19, along the three directions, the line profile is characterized by a complex, multi-peaked structure. This resembles the “Disturbed Disk” (DD) stage analysed in \citet{kohandel2020}, where the presence of multiple star-forming clumps of gas perturbs the main galaxy disk (see Fig. \ref{fig:E19_maps_int}). Furthermore, we remark that for many of the LOSs shown in Fig. \ref{fig:Lalpha_E31} the H$_{\rm \alpha}$ line shows a single peak shape, apart from E31, along the z and y directions, and E52 along the x direction. Along these last three LOSs, the H$_{\rm \alpha}$ profile is characterised by a double peak shape. This is consistent with what has been noted at the end of Sec. \ref{3Dintr}, namely that the gas velocity distribution along these directions resembles the dynamics of a rotating disk in the edge-on view\footnote{The spectral profile of a disk in the edge-on (face-on) view is characterised by a double (single) peak shape \citep[e.g.][]{elitzur2012}.}.

Independently on the shape of the H$_{\rm \alpha}$ line, it can be seen that the total flux (black solid line) always coincides with the flux arising from the star-forming regions (red dotted line), along any of the LOS analysed, and for all the events considered.  
 
This means that the merging MBHs in our simulation are not accreting efficiently enough to leave any signature in the synthetic H$_{\rm \alpha}$ line.

\subsection{Boosting the accretion rates and masses of the MBHs}\label{boost}
In the previous subsection, we concluded that the properties of our simulated systems are such that no signature of merging MBHs can be seen from the H$_{\rm \alpha}$ profile. To make this result more generic, we now ask ourselves the following question: {\it How efficiently two merging MBHs should accrete in order to be detectable?}

To answer this question, we re-simulate the H$_{\rm \alpha}$ line, keeping constant the contribution from the SFR, and artificially boosting the accretion rates and masses of the merging MBHs (both in the primary and in the secondary). For simplicity's sake, we restrict the analysis to a single event and to a single LOS. We choose E36 and the LOS along the x direction since, in this case, the H$_{\rm \alpha}$ line neither resembles a DD stage nor the dynamics of a rotating disk in the edge-on view.

The results of this experiment are shown in Fig. \ref{fig:Lalpha_E36_boost}: in the left column, we keep the masses of the MBHs involved in the merging of E36, and we vary their accretion rates from Eddington ($\lambda_{\rm EDD,p}=\lambda_{\rm EDD,s}=1$), up to 10$\times$ Eddington; in the right column, we repeat the same exercise in terms of the accretion rate, but we consider a primary MBH that is 10$\times$ more massive than the original one. While doing these calculations, we assume $L_{\rm bol}\propto \lambda_{\rm EDD}$, though we are aware of the sub-linear increasing of the luminosity with the accretion in super-Eddington regimes \citep[e.g.][]{madau2014}. We verify {\it a posteriori} that our assumption is conservative.

From Fig. \ref{fig:Lalpha_E36_boost}, it is clear (and obvious) that the more the MBHs are accreting and the more they are massive, the larger is the deviation from the original H$_{\rm \alpha}$ line. The point here is whether or not, even in these extreme cases, we would be able to infer the presence of the two accreting MBHs.

To understand this point, we adopt a procedure similar to the one described in \citet[][see also \citealt{carniani2023}]{Gallerani2018}, developed to infer the presence of outflowing gas from the shape of the [CII] (H$_{\rm \alpha}$) emission line. This method is based on the analysis of the residuals, as obtained by subtracting from the emission line data its best-fitting model. We thus fit our simulated H$_{\rm \alpha}$ lines both with a single Gaussian (SG) and a double Gaussian (DG) profile. The green and orange lines in Fig. \ref{fig:Lalpha_E36_boost} report the best fitting results, assuming an SG and a DG profile, respectively. In the bottom panel of each figure, we further report the residual of the best-fitting procedure, along with the random noise added to the synthetic spectrum. 

To measure the goodness of the fit, we apply the two-sample Kolmogorov–Smirnov (K-S) test  \citep{kolmogorov1933sulla,smirnov1948table} to our simulated data and we compute the K-S probability ($p_{KS}$). In particular, we apply the two-sample KS test to the cumulative distribution function (CDF) of the SG residual versus the CDF of the noise and to the CDF of the DG residual versus the CDF of the noise to test whether the residuals come from the same random distribution used to simulate the noise. We remind that, according to the K-S test, two samples are not drawn from the same underlying distribution if $p_{KS}<0.05$. We report the results of the K-S test in the bottom panels of each figure, along with the best-fit parameters, where the amplitudes of the Gaussians are in units of $10^{-19}\rm erg~s^{-1}~{cm}^{-2}~A^{-1}$.  

We find that, even in the case $\lambda_{\rm EDD,p}=\lambda_{\rm EDD,s}=1$ (top, left panel), the H$_{\rm \alpha}$ line can be well fitted with an SG profile. In all the other cases, the extreme accretion rates considered lead the profile to deviate in a statistically significant way from an SG profile ($p_{KS,SG}<0.05$). However, we also find that, in all the cases, a DG profile is enough to provide a good fit to the synthetic spectra ($p_{KS,DG}>0.05$).

To summarise, the presence of two accreting MBHs would be difficult to infer even in very extreme (i.e. rare) circumstances ($\lambda_{\rm EDD}\sim 5-10$). We underline that this is true even if, with the assumption $L_{\rm bol}\propto \lambda_{\rm EDD}$, we are maximising the effects of the multiple accreting MBHs on the shape of the H$_{\rm \alpha}$ line. We thus conclude that the combined detection of GW and EM signals from $z\gtrsim 6$ MBHs is challenging (if not impossible) not only because of the poor sky-localization ($\sim$10~$\rm deg^2$) provided by LISA but also because the loudest GW emitters ($M_{\rm BH}\sim 10^{5-6}~\rm M_{\odot}$) are not massive enough to leave significant signatures (e.g. extended wings) in the emission lines arising from the broad line region.

\section{Summary and discussion}\label{dis}

 In this work, we adopted a zoom-in cosmological hydrodynamical simulation of galaxy formation and black hole (BH) co-evolution, based on the GADGET-3 code, zoomed-in on a $M_h \sim 10^{12}~\rm M_{\odot}$ dark matter halo at $z = 6$, which hosts a fast accreting ($\dot{M}\sim 35~\rm M_{\odot}~yr^{-1}$) super-massive black hole (SMBH, $M_{\rm BH}\sim 10^{9}~\rm M_{\odot}$) and a 
 star-forming galaxy (SFR $\sim 100~\rm M_{\odot}~yr^{-1}$). 
 
 Following the SMBH formation backward in time, we have identified the merging events that concurred to its formation and we have focused our analysis on the ones that are detectable with LISA and that, after considering the effect of delay due to dynamical friction in the MBH coalescence, are still occurring at $z\gtrsim 6$. These arise from the coalescence of massive black holes (MBHs) with masses $M_{\rm BH}\sim 10^{5-6}~\rm M_{\odot}$. 
 
 We have then investigated the intrinsic properties of the host galaxies of these LISA detectable events (LDEs), finding the following typical properties: BH accretion rate $\dot{M}\sim 0.01~\rm M_{\odot}~yr^{-1}$, star formation rate SFR$\sim 10-40~\msun~\rm yr^{-1}$, metallicity $Z\sim 0.4-0.6~Z_{\odot}$, dust mass  $M_{\rm dust} \sim 10^5-10^6~\msun$, stellar mass $M_{\rm stellar} \sim 10^9-10^{10}~\msun$, and gas mass $M_{\rm gas} \sim 10^9-10^{10}~\msun$.
 
 Among these LDEs, we have selected those that, based on their intrinsic properties are expected to be bright in one or more electromagnetic (EM) bands (e.g. rest-frame X-ray, UV, and FIR). We find that $\sim$20-30\% of the LDEs are also detectable with EM telescopes. 
 
 We have post-processed these events with dust radiative transfer calculations to make accurate predictions about their spectral energy distributions (SEDs) and continuum maps in the JWST to ALMA wavelength range. By comparing the spectra arising from galaxies hosting the merging MBHs with those arising from AGN powered by single accreting BHs, we have found that it will be impossible to identify an LDE from the continuum SEDs because of the absence of specific imprints from the merging MBHs. 
 
 Finally, we have computed the profile of the H$_{\rm \alpha}$ line arising from LDEs, taking into account both the contribution from their star-forming regions and the accreting MBHs. We find that even in the extreme case of both MBHs accreting at super-Eddington rates the shape of the H$_{\rm \alpha}$ line does not deviate significantly from the one arising from star formation only. 
 
 We conclude that the combined detection of GW and EM signals from $z\gtrsim 6$ MBHs is challenging (if not impossible) not only because of the poor sky-localization ($\sim$10~$\rm deg^2$) provided by LISA, but also because the loudest GW emitters ($M_{\rm BH}\sim 10^{5-6}~\rm M_{\odot}$) are not massive enough to leave significant signatures (e.g. extended wings) in the emission lines arising from the broad line region.

\section*{Acknowledgements}

SG acknowledges support from the ASI-INAF n. 2018-31-HH.0 grant and PRIN-MIUR 2017. MV is supported by the Italian Research Center on High Performance Computing, Big Data and Quantum Computing (ICSC), project funded by European Union - NextGenerationEU - and National Recovery and Resilience Plan (NRRP) - Mission 4 Component 2, within the activities of Spoke 3, Astrophysics and Cosmos Observations, and by the INFN Indark Grant. 


\bibliographystyle{aa}
\bibliography{references} 
\appendix 

\section{3D representation of intrinsic properties for the other events}

In Fig.~\ref{fig:E31_maps_int} we showed the intrinsic properties (gas density, SFR, gas metallicity and gas velocity) relative to E31. In this Section, we show the same properties for events E19, E36, E52.

\begin{figure*}
	\centering
 \includegraphics[width=1.\textwidth] {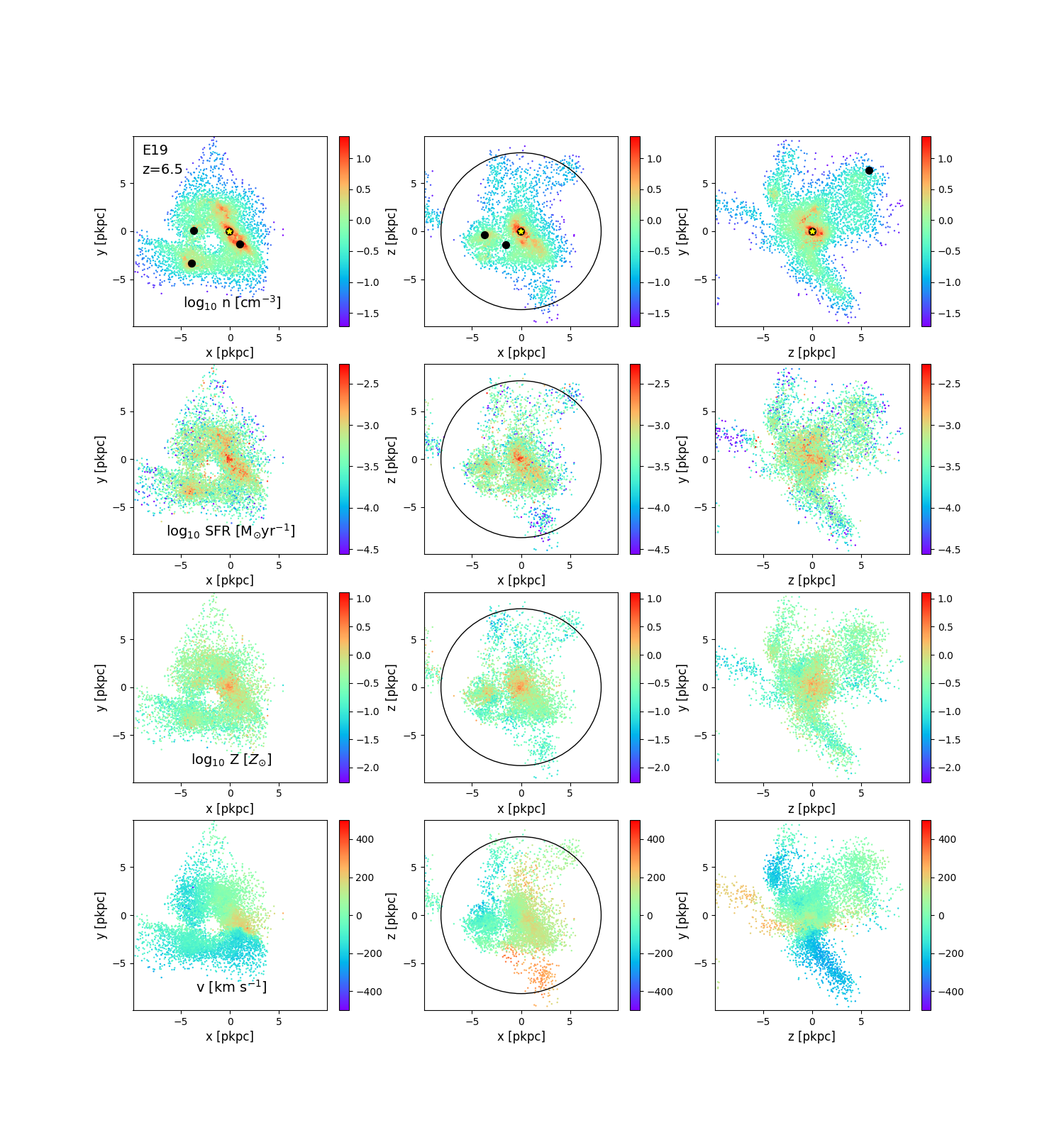}
    \caption{Same as in Fig. \ref{fig:E31_maps_int}, but for E19.}
    \label{fig:E19_maps_int}
\end{figure*}
\begin{figure*}
	\centering
 \includegraphics[width=1.\textwidth] {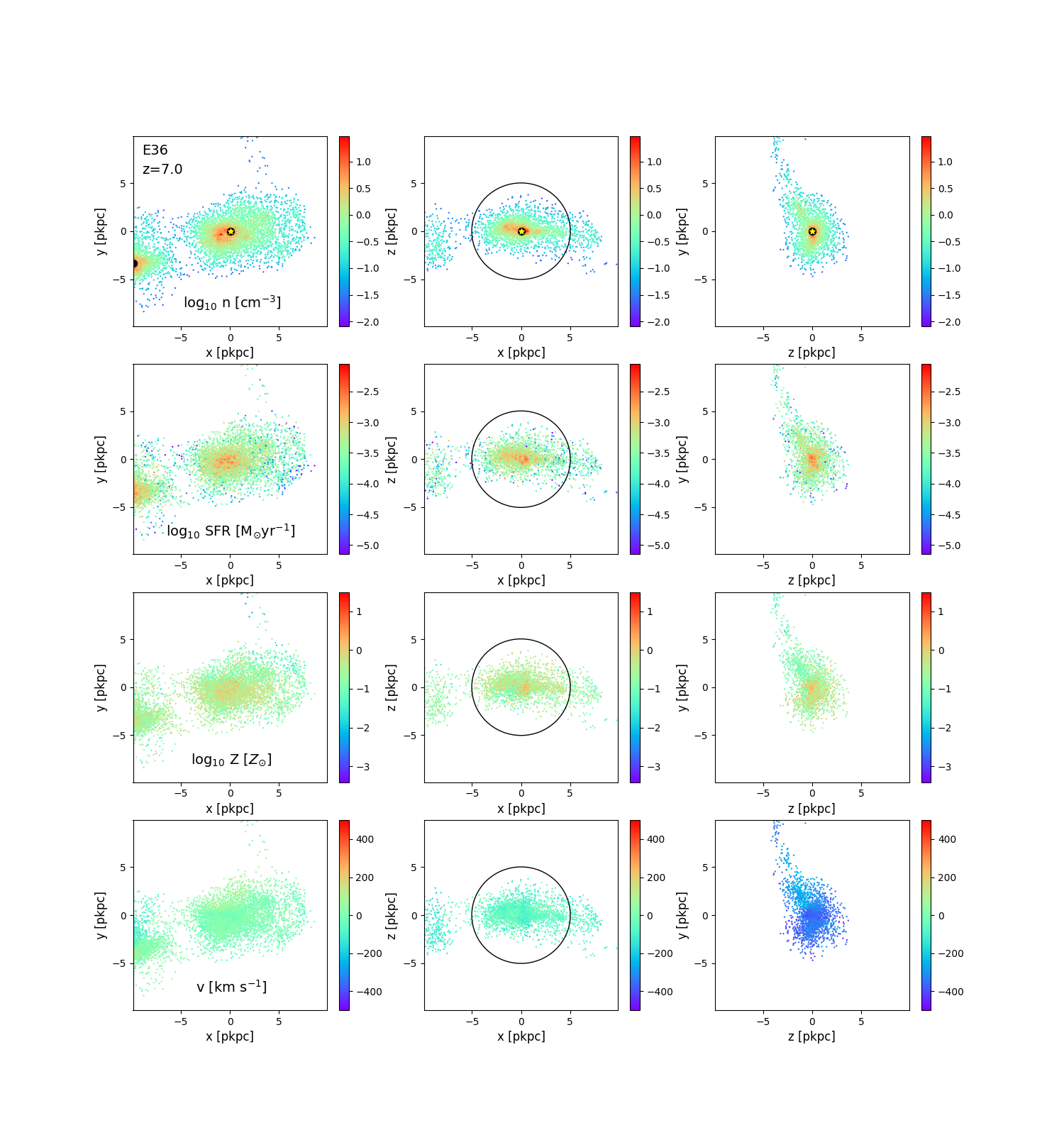}
    \caption{Same as in Fig. \ref{fig:E31_maps_int}, but for E36.}
    \label{fig:E36_maps_int}
\end{figure*}
\begin{figure*}
	\centering
 \includegraphics[width=1.\textwidth] {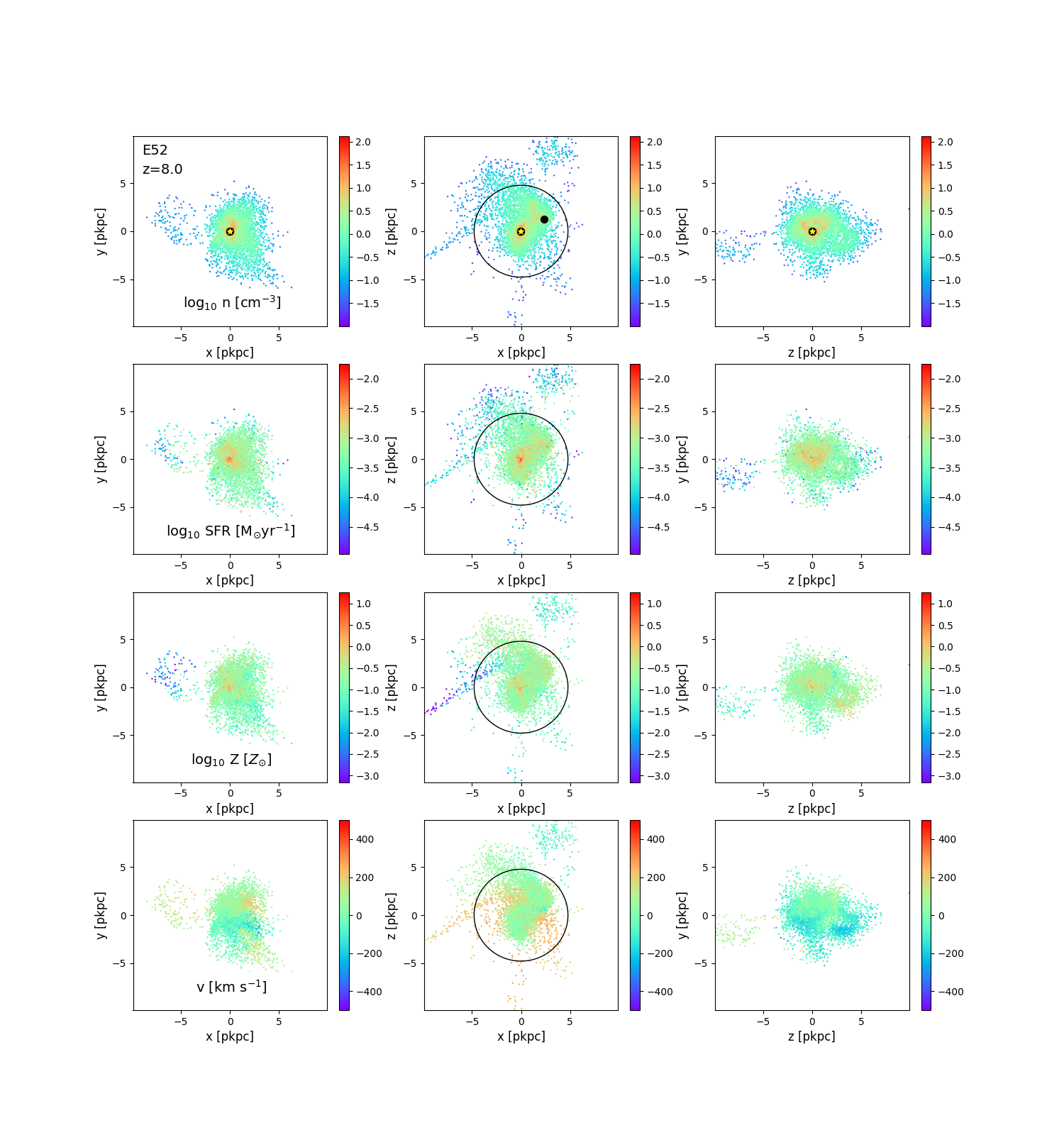}
    \caption{Same as in Fig. \ref{fig:E31_maps_int}, but for E52.}
    \label{fig:E52_maps_int}
\end{figure*}

\section{Observable properties}\label{MWL}
\subsection{UV and X-ray Luminosity}
We compute the UV luminosity

considering the contributions from both stars and accreting MBHs:
\begin{equation}
    L_{\rm UV}=L_{\rm UV,*}+L_{\rm UV,AGN}.
\end{equation}
For what concerns stars, we adopt the SFR-$L_{\rm UV,*}$ relation derived by \citet{Murphy_2011}:
\begin{equation}
\left(\frac{\rm SFR_{UV}}{M_{\sun}~{\rm yr^{-1}}}\right) = 4.42\times10^{-44}\left(\frac{L_{\rm UV,*}}{\rm erg~s^{-1}}\right).
\end{equation}
The AGN UV luminosity is instead computed as follows:
\begin{equation}
    L_{\rm UV,AGN}=f_{\rm UV} L_{\rm bol} ,
\end{equation}
where the bolometric luminosity $L_{\rm bol}$ is related
to the accretion rate of the MBH, $L_{\rm bol}= \epsilon_ r \dot{M} c^2$, $\epsilon_ r$ is the radiative efficiency, and $f_{\rm UV}$ represents the UV bolometric correction parametrized as in \citet{Hopkins_2007}: 
 \begin{equation}
    \dfrac{L_{\rm bol}}{L_{\rm band}}= c_1 \Big(\dfrac{L_{\rm bol}}{10^{10}L_{\odot}}\Big)^{k_1} + c_2 \Big(\dfrac{L_{\rm bol}}{10^{10}L_{\odot}}\Big)^{k_2},
\end{equation}
 where, $c_1=1.862$, $k_1=-0.361$, $c_2=4.870$, $k_2=-0.0063$ \citep{Shen_2020}. 

 At $\lambda$ = 1450 $\angstrom$, we can relate the AGN UV luminosity to the emitted UV luminosity density as:
 \begin{equation}
   \nu L_{\rm \nu} = L_{\rm UV,AGN}.
\end{equation}
 The observed UV flux density can be then expressed as:
 \begin{equation}
     f_{\nu} = \frac{L_{\rm \nu}}{4 \pi d_L^2},
 \end{equation}
 where $d_L$ is the luminosity distance of the source.
 The observed(apparent) UV magnitude can be derived from the flux density \citep{Oke_1983}:
 \begin{equation}
   m_{\rm AB} = -2.5~ log_{10} \frac{f_{\nu}}{\rm ergs~ s^{-1} cm^{-2}Hz^{-1}}-48.6,
\end{equation}
and the apparent ($m_{\rm AB}$) and absolute magnitude ($M_{\rm AB}$) are related as:
\begin{equation}
   M_{\rm AB} = m_{\rm AB}-\mu.
\end{equation}
 where $\mu$ is the distance modulus.

Similarly, we compute the X-ray luminosity

as $L_{\rm X}=f_{\rm X} L_{\rm bol}$, with the following parameters for the bolometric correction: $c_1=4.073$, $k_1=-0.026$, $c_2=12.6$, $k_2=0.278$ \citep{Hopkins_2007}. These bolometric corrections well reproduce the X-ray SED template proposed by \citet{Shen_2020}.
\subsection{Far-infrared luminosity}
The far-infrared luminosity associated with dust emission in the optically-thin regime can be computed as follows \citep{Carniani_2019}: 
\begin{equation}
    {L_{\rm FIR}}= 4\pi{M_{\rm dust}}\int k_\nu{B_{\rm \nu}}({T_{\rm dust}})d\nu ,
\end{equation}
where $B_{\rm \nu}({T_{\rm dust}})$ is the Planck function associated to the dust component at temperature $T_{\rm dust}$:
\begin{equation}
    {B_{\rm \nu}}({T_{\rm dust}})=\frac{2h\nu^3}{c^2}\frac{1}{e^{\frac{h\nu}{k_BT_{\rm dust}}}-1} ,
\end{equation}
and
\begin{equation}
    k_{\nu}=k_{0}\Big(\dfrac{\nu}{\nu_0}\Big)^{\beta} ,
\end{equation}
where we assume the dust emissivity index $\beta = 2.2$, and the mass absorption coefficient $k_0=34.7$ at a frequency $\nu_0$ that corresponds to $\lambda=100~\rm\mu m$ (\citealt{doi:10.1146/annurev.astro.41.011802.094840,Draine2004}). These parameters fit the dust emissivity in the far-infrared band $k_\nu$ of the Small Magenallic Cloud. The dust mass is given by $M_{\rm dust} = f_{\rm d} \times M_{\rm metals}$, with $f_{\rm d}$ = 0.1 and $M_{\rm metals}=Z \times M_{\rm gas}$, as described in Sec.~\ref{sec:dust_properties}.
\subsection{[CII] Luminosity}
To compute the [CII] luminosity $L_{\rm CII}$ we adopt the following fitting formula proposed by \citet{Yue_2013} and based on the ISM sub-grid models by \citet{Vallini_2013,Vallini_2015}:
\begin{align}
{\rm log}(L_{\rm CII}) &= 7.0 + 1.2\times{\rm log(SFR)} + 0.021\times{\rm log(Z)} \nonumber \\
&+ 0.012\times{\rm log(SFR)log(Z)} - 0.74\times{\rm log^2(Z)},
\label{LCII}
\end{align}
where $L_{\rm CII}$, SFR, and $Z$ are given in units of $\rm L_\odot$, $M_\odot$yr$^{-1}$ and $Z_\odot$, respectively.
\begin{table*}
\caption{Intrinsic properties of GW host galaxies. In this table, for each merging event (first column) associated with a host galaxy in the \AGNfiducial{} case, we report the redshift of the merger (second column), the distance between the primary BH and the associated galaxy (third column), the masses of the merging BHs (fourth and fifth column), the accretion rate (in $\msun \rm yr^{-1}$) of the merged MBH (sixth column), the star formation rate (in $\msun \rm yr^{-1}$, seventh column), the stellar mass (eighth column), the dust mass (ninth column) and the metallicity (tenth column) in solar units.}
\label{intrinsicprops}
\centering
\resizebox{\textwidth}{!}{%
\begin{tabular}{|c|c|c|c|c|c|c|c|c|c|}
\hline
\textit{\textbf{Event Number}} & \textit{\textbf{Redshift}} & \textit{\textbf{\begin{tabular}[c]{@{}c@{}}Distance between galaxy\\ and BH (physical kpc)\end{tabular}}} & \textit{\textbf{\begin{tabular}[c]{@{}c@{}}$M_1$\\ ($\rm M_{\odot}$)\end{tabular}}} & \textit{\textbf{\begin{tabular}[c]{@{}c@{}}$M_2$\\ ($\rm M_{\odot}$)\end{tabular}}} & \textit{\textbf{\begin{tabular}[c]{@{}c@{}}$\dot{M}$\\ ($\rm M_{\odot} \rm yr^{-1}$)\end{tabular}}} & \textit{\textbf{\begin{tabular}[c]{@{}c@{}}SFR\\ ($\rm M_{\odot} \rm yr^{-1}$)\end{tabular}}} & \textit{\textbf{\begin{tabular}[c]{@{}c@{}}$M_{\rm stellar}$\\ ($\rm M_{\odot}$)\end{tabular}}} & \textit{\textbf{\begin{tabular}[c]{@{}c@{}}$M_{\rm dust}$\\ ($\rm M_{\odot}$)\end{tabular}}} & \textit{\textbf{\begin{tabular}[c]{@{}c@{}}$Metallicity$\\ ($Z_{\odot}$)\end{tabular}}} \\ \hline
1 & 6 & 0.48 & 1.31$\times 10^6$ & 1.02$\times 10^5$ & 0.012 & 13.1 & $ 2.02 \times 10^{9} $ & $ 1.7 \times 10^{6} $ & 0.88 \\ \hline
2 & 6 & 0.08 & 6.67$\times 10^8$ & 1$\times 10^5$ & 37 & 166 & $2.53 \times 10^{10} $ & $ 2.5 \times 10^{7} $ & 0.62 \\ \hline
3 & 6.09 & 0.51 & 1.01$\times 10^{5}$ & 1$\times 10^{5}$ & 1.58$\times 10^{-5}$ & 0.03 & 1.06$\times 10^{7}$ & 9.43$\times 10^{3}$ & 0.13 \\ \hline
4 & 6.12 & 0.59 & $ 3.71 \times 10^{8} $ & $ 1.03 \times 10^{5} $ & 11.3 & 142 & $ 2.41 \times 10^{10} $ & $ 2.1 \times 10^{7} $ & 0.61 \\ \hline
5 & 6.13 & 3.57 & 3.82$\times 10^{5}$ & 1$\times 10^{5}$ & 2.9$\times 10^{-2}$ & 21.9 & 5.55$\times 10^{9}$ & 6.76$\times 10^{6}$ & 0.72 \\ \hline
6 & 6.17 & 4.36 & 2.17$\times 10^{5}$ & 1.01$\times 10^{5}$ & 2.9 & 116 & 2.1$\times 10^{10}$ & 1.88$\times 10^{7}$ & 0.59 \\ \hline
7 & 6.19 & 0.89 & 9.86$\times 10^{5}$ & 1$\times 10^{5}$ & 5.3$\times 10^{-3}$ & 11.3 & 2.02$\times 10^{9}$ & 2.16$\times 10^{6}$ & 0.57 \\ \hline
8 & 6.22 & 1.55 & 7.58$\times 10^{5}$ & 1.04$\times 10^{5}$ & 1.37$\times 10^{-3}$ & 9.84 & 1.18$\times 10^{9}$ & 1.31$\times 10^{6}$ & 0.75 \\ \hline
9 & 6.22 & 1.84 & 2.03$\times 10^{5}$ & 1.01$\times 10^{5}$ & 2.83$\times 10^{-2}$ & 21.2 & 5.33$\times 10^{9}$ & 6.65$\times 10^{6}$ & 0.73 \\ \hline
10 & 6.29 & 0.82 & 8.07$\times 10^{5}$ & 1.01$\times 10^{5}$ & 5.18$\times 10^{-3}$ & 10.9 & 1.95$\times 10^{9}$ & 2.14$\times 10^{6}$ & 0.53 \\ \hline
11 & 6.29 & 0.22 & 2.37$\times 10^{8}$ & 1.0$\times 10^{5}$ & 3.42 & 83.7 & 1.71$\times 10^{10}$ & 1.46$\times 10^{7}$ & 0.56 \\ \hline
12 & 6.3 & 0.34 & 4.82$\times 10^{5}$ & 1.02$\times 10^{5}$ & 1.07$\times 10^{-2}$ & 8.19 & 1.15$\times 10^{9}$ & 1.34$\times 10^{6}$ & 0.61 \\ \hline
13 & 6.31 & 1.08 & 1.01$\times 10^{5}$ & 1$\times 10^{5}$ & 1.95$\times 10^{-5}$ & 0.11 & 1.49$\times 10^{7}$ & 1.71$\times 10^{4}$ & 0.11 \\ \hline
14 & 6.36 & 0.75 & 2.2$\times 10^{8}$ & 1.0$\times 10^{5}$ & 1.25 & 59.6 & 1.2$\times 10^{10}$ & 1.1$\times 10^{7}$ & 0.52 \\ \hline
15 & 6.4 & 0.13 & 1.09$\times 10^{5}$ & 1.01$\times 10^{5}$ & 1.82$\times 10^{-4}$ & 1.12 & 1.71$\times 10^{8}$ & 2.06$\times 10^{5}$ & 0.29 \\ \hline
16 & 6.41 & 1.16 & 1.47$\times 10^{8}$ & 1.01$\times 10^{5}$ & 0.78 & 27.1 & 9.96$\times 10^{9}$ & 6.46$\times 10^{6}$ & 0.61 \\ \hline
17 & 6.42 & 2.23 & 1.03$\times 10^{5}$ & 1$\times 10^{5}$ & 2.27$\times 10^{-3}$ & 7.72 & 7.55$\times 10^{8}$ & 1.02$\times 10^{6}$ & 0.45 \\ \hline
18 & 6.5 & 0.04 & 1.39$\times 10^{8}$ & 7.84$\times 10^{5}$ & 0.72 & 29.6 & 9.63$\times 10^{9}$ & 6.46$\times 10^{6}$ & 0.60 \\ \hline
19* & 6.5 & 0.17 & 1.25$\times 10^{6}$ & 2.01$\times 10^{5}$ & 9.84$\times 10^{-3}$ & 31.08 & 4.8$\times 10^{9}$ & 5.82$\times 10^{6}$ & 0.53 \\ \hline
20 & 6.51 & 0.67 & 3.78$\times 10^{5}$ & 1.03$\times 10^{5}$ & 4.55$\times 10^{-3}$ & 6.96 & 1.27$\times 10^{9}$ & 1.26$\times 10^{6}$ & 0.63 \\ \hline
21 & 6.51 & 0.79 & 4.33$\times 10^{5}$ & 1$\times 10^{5}$ & 5.89$\times 10^{-3}$ & 2.06 & 5.98$\times 10^{8}$ & 5.17$\times 10^{5}$ & 0.58 \\ \hline
22 & 6.51 & 0.88 & 6.66$\times 10^{6}$ & 1.0$\times 10^{5}$ & 8.32$\times 10^{-2}$ & 8.5 & 1.11$\times 10^{9}$ & 7.49$\times 10^{5}$ & 0.72 \\ \hline
23 & 6.55 & 1.23 & 1.65$\times 10^{8}$ & 1.01$\times 10^{5}$ & 0.77 & 50.46 & 1.06$\times 10^{10}$ & 9.29$\times 10^{6}$ & 0.52 \\ \hline
24 & 6.6 & 0.31 & 1.04$\times 10^{5}$ & 1.01$\times 10^{5}$ & 9.49$\times 10^{-4}$ & 7.31 & 6.83$\times 10^{8}$ & 1.02$\times 10^{6}$ & 0.35 \\ \hline
25 & 6.61 & 0.28 & 1.45$\times 10^{8}$ & 1.0$\times 10^{5}$ & 1.72 & 39.33 & 7.88$\times 10^{9}$ & 7.05$\times 10^{6}$ & 0.46 \\ \hline
26 & 6.68 & 0.71 & 1.02$\times 10^{8}$ & 9.05$\times 10^{6}$ & 3.93 & 33.28 & 6.95$\times 10^{9}$ & 5.46$\times 10^{6}$ & 0.46 \\ \hline
27 & 6.69 & 0.45 & 1.02$\times 10^{5}$ & 1.0$\times 10^{5}$ & 4.72$\times 10^{-5}$ & 1.14 & 1.83$\times 10^{8}$ & 2.65$\times 10^{5}$ & 0.51 \\ \hline
28 & 6.7 & 0.07 & 1.05$\times 10^{5}$ & 1.01$\times 10^{5}$ & 3.02$\times 10^{-5}$ & 1.02 & 1.71$\times 10^{8}$ & 2.17$\times 10^{5}$ & 0.36 \\ \hline
29 & 6.76 & 1.4 & 6.41$\times 10^{7}$ & 1.0$\times 10^{5}$ & 9.84$\times 10^{-1}$ & 33.21 & 6.38$\times 10^{9}$ & 4.99$\times 10^{6}$ & 0.47 \\ \hline
30 & 6.79 & 0.40 & 1.02$\times 10^{5}$ & 1.0$\times 10^{5}$ & 2.9$\times 10^{-6}$ & 0.31 & 8.77$\times 10^{7}$ & 9.02$\times 10^{4}$ & 0.84 \\ \hline
31* & 6.8 & 0.27 & 5.24$\times 10^{5}$ & 1.01$\times 10^{5}$ & 1.9$\times 10^{-2}$ & 34.7 & 2.88$\times 10^{9}$ & 3.87$\times 10^{6}$ & 0.41 \\ \hline
32 & 6.91 & 0.68 & $ 2.4 \times 10^{5} $ & $ 2.1 \times 10^{5} $ & 0.0031 & 27.48 & $ 2.4 \times 10^{9} $ & $ 3.4 \times 10^{6} $ & 0.38 \\ \hline
33 & 6.92 & 0.88 & $ 1 \times 10^{5} $ & $ 1 \times 10^{5} $ & $ 3 \times 10^{-5} $ & 1.86 & $ 2.4 \times 10^{8} $ & $ 3.2 \times 10^{5} $ & 0.41 \\ \hline
34 & 6.96 & 2.71 & $ 4.5 \times 10^{5} $ & $ 1 \times 10^{5} $ & 0.0045 & 5.55 & $ 7.5 \times 10^{8} $ & $ 7.1 \times 10^{5} $ & 0.48 \\ \hline
35 & 6.98 & 1.80 & $ 1.2 \times 10^{5} $ & $ 1 \times 10^{5} $ & 0.0005 & 2.98 & $ 4.2 \times 10^{8} $ & $ 4.7 \times 10^{5} $ & 0.38 \\ \hline
36* & 7.01 & 0.65 & 2.51$\times 10^{5}$ & 1.02$\times 10^{5}$ & 0.0026 & 6.77 & $ 8.4 \times 10^{8} $ & $ 1.2 \times 10^{6} $ & 0.44 \\ \hline
37 & 7.04 & 2.49 & $ 9.9 \times 10^{7} $ & $ 1 \times 10^{5} $ & 0.38 & 21.97 & $ 7.9 \times 10^{9} $ & $ 5.2 \times 10^{6} $ & 0.65 \\ \hline
38 & 7.2 & 0.15 & 8.67$\times 10^{7}$ & 1.0$\times 10^{5}$ & 0.89 & 26.35 & $ 6.6 \times 10^{9} $ & $ 4.9 \times 10^{6} $ & 0.59 \\ \hline
39 & 7.24 & 1.32 & $ 2.8 \times 10^{5} $ & $ 1 \times 10^{5} $ & 0.021 & 5.68 & $ 5.8 \times 10^{8} $ & $ 6.7 \times 10^{5} $ & 0.38 \\ \hline
40 & 7.24 & 2.19 & $ 1.2 \times 10^{6} $ & $ 1 \times 10^{5} $ & 0.013 & 16.13 & $1.5 \times 10^{9} $ & $ 1.4 \times 10^{6} $ & 0.46 \\ \hline
41 & 7.32 & 1.33 & $ 1 \times 10^{5} $ & $ 1 \times 10^{5} $ & $ 1.9 \times 10^{-6} $ & 5.56 & $ 3.2 \times 10^{8} $ & $ 5.1 \times 10^{5} $ & 0.33 \\ \hline
42 & 7.36 & 1.28 & $ 3.6 \times 10^{6} $ & $ 1 \times 10^{5} $ & 0.024 & 21.32 & $ 3.2 \times 10^{9} $ & $ 2.3 \times 10^{6} $ & 0.44 \\ \hline
43 & 7.4 & 0.09 & 1.02$\times 10^{5}$ & 1.01$\times 10^{5}$ & 0.00019 & 5.17 & $ 3.5 \times 10^{8} $ & $ 5 \times 10^{5} $ & 0.25 \\ \hline
44 & 7.48 & 0.61 & 1.03$\times 10^{5}$ & 1.02$\times 10^{5}$ & $ 1.8 \times 10^{-5} $ & 6.76 & $ 5.3 \times 10^{8} $ & $ 8.2 \times 10^{5} $ & 0.28 \\ \hline
45 & 7.48 & 0.72 & 1.01$\times 10^{5}$ & 1.01$\times 10^{5}$ & $ 1.2 \times 10^{-5} $ & 0.25 & $ 3.7 \times 10^{7} $ & $ 4.6 \times 10^{4} $ & 0.18 \\ \hline
46 & 7.53 & 1.41 & $ 6.5 \times 10^{5} $ & $ 1 \times 10^{5} $ & 0.0079 & 12.99 & $ 1.1 \times 10^{9} $ & $ 1.2 \times 10^{6} $ & 0.39 \\ \hline
47 & 7.58 & 0.52 & 1.04$\times 10^{5}$ & 1.01$\times 10^{5}$ & $1.3\times 10^{-4}$ & 1.98 & $ 1.7 \times 10^{8} $ & $ 2.5 \times 10^{5} $ & 0.28 \\ \hline
48 & 7.62 & 1.05 & $ 5.3 \times 10^{7} $ & $ 1 \times 10^{5} $ & 0.44 & 31.52 & $ 6.4 \times 10^{9} $ & $ 4.6 \times 10^{6} $ & 0.63 \\ \hline
49 & 7.75 & 1.18 & $ 1 \times 10^{5} $ & $ 1 \times 10^{5} $ & $1.7\times 10^{-4}$ & 4.46 & $ 3.3 \times 10^{8} $ & $ 5 \times 10^{5} $ & 0.28 \\ \hline
50 & 7.9 & 0.19 & 2.87$\times 10^{7}$ & 9.57$\times 10^{5}$ & 1 & 31.32 & $ 4.7 \times 10^{9} $ & $ 4 \times 10^{6} $ & 0.6 \\ \hline
51 & 7.98 & 0.79 & 2.06$\times 10^{5}$ & 1.0$\times 10^{5}$ & 4.6$\times 10^{-4}$ & 1.13 & 1.3$\times 10^{8}$ & 1.8$\times 10^{5}$ & 0.30 \\ \hline
52* & 8 & 0.01 & 1.49$\times 10^{6}$ & 1.02$\times 10^{5}$ & 0.13 & 34.7 & $ 2.1 \times 10^{9} $ & $ 2.3 \times 10^{6} $ & 0.36 \\ \hline
53 & 8.01 & 0.39 & 1.22$\times 10^{5}$ & 1.01$\times 10^{5}$ & $ 7.4 \times 10^{-5} $ & 2.22 & $ 3.2 \times 10^{8} $ & $ 3.4 \times 10^{5} $ & 0.37 \\ \hline
54 & 8.02 & 0.52 & 1.02$\times 10^{5}$ & 1.01$\times 10^{5}$ & $ 8.3 \times 10^{-5} $ & 3.13 & $ 2.6 \times 10^{8} $ & $ 3.5 \times 10^{5} $ & 0.25 \\ \hline
55 & 8.51 & 0.29 & 2.18$\times 10^{5}$ & 1.01$\times 10^{5}$ & 8.60$\times 10^{-4}$ & 4.12 & 2.35$\times 10^{8}$ & 3.80$\times 10^{5}$ & 0.25 \\ \hline
56 & 8.99 & 0.07 & 1.06$\times 10^{5}$ & 1.01$\times 10^{5}$ & 2.02$\times 10^{-4}$ & 5.43 & 2.8$\times 10^{8}$ & 3.6$\times 10^{5}$ & 0.25 \\ \hline
\end{tabular}%
}
\end{table*}
\begin{table*}
\caption{Observable properties of GW host galaxies. In this table, for each GW event (first column) associated with a host galaxy in the \AGNfiducial{} case, we report the redshift of the merger (second column), the distance between the primary BH and the associated galaxy (third column), the chirp mass (fourth column), the X-ray (fifth column), [CII] (sixth column) and FIR (seventh column) luminosities in solar units, the apparent and absolute UV magnitudes (eighth and ninth column), the SNR (tenth column) and the angular resolution (eleventh column). We show the cases with either significant SNR or significant brightness in different EM bands.}
\label{observable}
\centering
\resizebox{\textwidth}{!}{%
\begin{tabular}{|c|c|c|c|c|c|c|c|c|c|c|}
\hline
\textit{\textbf{\begin{tabular}[c]{@{}c@{}}Event\\ Number\end{tabular}}} & \textit{\textbf{Redshift}} & \textit{\textbf{\begin{tabular}[c]{@{}c@{}}Distance between galaxy\\ and BH (physical kpc)\end{tabular}}} & \textit{\textbf{\begin{tabular}[c]{@{}c@{}}${\mathcal M_c}$\\ (M/$M_{\odot}$)\end{tabular}}} & \textit{\textbf{\begin{tabular}[c]{@{}c@{}}$ L_x$\\ (ergs/s)\end{tabular}}} & \textit{\textbf{\begin{tabular}[c]{@{}c@{}}$ L_{\rm CII}$\\ (L/$L_{\odot}$)\end{tabular}}} & \textit{\textbf{\begin{tabular}[c]{@{}c@{}}$L_{\rm FIR}$\\ (L/$L_{\odot}$)\end{tabular}}} & \textit{\textbf{\begin{tabular}[c]{@{}c@{}}Apparent\\ Magnitude\end{tabular}}} & \textit{\textbf{$M_{\rm ab}$}} & \textit{\textbf{SNR}} & \textit{\textbf{\begin{tabular}[c]{@{}c@{}}Angular Resolution\\ $\Omega$\\ ($Deg^2$)\end{tabular}}} \\ \hline
1 & 6 & 0.48 & $ 1.9 \times 10^{6} $ & $ 1.4 \times 10^{42} $ & $ 2.1 \times 10^{8}$ & $3 \times 10^{12} $ & 28.01 & -20.85 & 24.12 & 859.33 \\ \hline
2 & 6 & 0.08 & $ 2.4 \times 10^{7} $ & $ 6.2 \times 10^{44} $ & $ 3.9 \times 10^{9} $ & $ 4.4 \times 10^{13} $ & 23.26 & -25.60 & \begin{tabular}[c]{@{}c@{}}Not Detectable\end{tabular} & Not Applicable \\ \hline
3 & 6.09 & 0.51 & $ 6.2 \times 10^{5} $ & $ 4.1 \times 10^{39} $ & $ 3.5 \times 10^{4} $ & $ 1.7 \times 10^{10} $ & 34.46 & -14.44 & 164.27 & 18.53 \\ \hline
4 & 6.12 & 0.59 & $ 1.9 \times 10^{7} $ & $ 2.6 \times 10^{44} $ & $ 3.2 \times 10^{9} $ & $ 3.8 \times 10^{13} $ & 24.34 & -24.57 & \begin{tabular}[c]{@{}c@{}}Not Detectable\end{tabular} & Not Applicable \\ \hline
5 & 6.13 & 3.57 & $ 1.2 \times 10^{6} $ & $ 2.8 \times 10^{42} $ & $ 3.7 \times 10^{8} $ & $ 1.2 \times 10^{13} $ & 27.48 & -21.42 & 89.62 & 62.25 \\ \hline
6 & 6.17 & 4.36 & $ 9.1 \times 10^{5} $ & $ 9.5 \times 10^{43} $ & $ 2.5 \times 10^{9} $ & $ 3.3 \times 10^{13} $ & 25.25 & -23.69 & 129.01 & 30.04 \\ \hline
7 & 6.19 & 0.89 & $ 1.8 \times 10^{6} $ & $ 7.1 \times 10^{41} $ & $ 1.5 \times 10^{8} $ & $ 3.8 \times 10^{12} $ & 28.27 & -20.68 & 32.90 & 461.85 \\ \hline
8 & 6.22 & 1.55 & $ 1.6 \times 10^{6} $ & $ 1.5 \times 10^{42} $ & $ 1.4 \times 10^{8} $ & $ 2.3 \times 10^{12} $ & 28.4 & -20.55 & 45.29 & 243.81 \\ \hline
9 & 6.22 & 1.84 & $ 8.9 \times 10^{5} $ & $ 2.8 \times 10^{42} $ & $ 3.6 \times 10^{8} $ & $ 1.2 \times 10^{13} $ & 27.56 & -21.38 & 131.39 & 28.96 \\ \hline
10 & 6.29 & 0.82 & $ 1.7 \times 10^{6} $ & $ 7 \times 10^{41} $ & $ 1.4 \times 10^{8} $ & $ 3.8 \times 10^{12} $ & 28.35 & -20.64 & 40.86 & 299.42 \\ \hline
11 & 6.29 & 0.22 & $ 1.6 \times 10^{7} $ & $ 1.1 \times 10^{44} $ & $ 1.6 \times 10^{9} $ & $ 2.6 \times 10^{13} $ & 25.42 & -23.56 & \begin{tabular}[c]{@{}c@{}}Not Detectable\end{tabular} & Not Applicable \\ \hline
12 & 6.3 & 0.34 & $ 1.3 \times 10^{6} $ & $ 1.3 \times 10^{42} $ & $ 1 \times 10^{8} $ & $ 2.4 \times 10^{12} $ & 28.64 & -20.35 & 69.82 & 102.57 \\ \hline
13 & 6.31 & 1.08 & $ 6.4 \times 10^{5} $ & $ 5 \times 10^{39} $ & $ 1.2 \times 10^{5} $ & $ 3 \times 10^{10} $ & 33.31 & -15.67 & 156.91 & 20.31 \\ \hline
14 & 6.36 & 0.75 & $ 1.6 \times 10^{7} $ & $ 5 \times 10^{43} $ & $ 1 \times 10^{9} $ & $ 1.9 \times 10^{13} $ & 26.13 & -22.89 & \begin{tabular}[c]{@{}c@{}}Not Detectable\end{tabular} & Not Applicable \\ \hline
15 & 6.4 & 0.13 & $ 6.8 \times 10^{5} $ & $ 3.9 \times 10^{40} $ & $ 5.7 \times 10^{6} $ & $ 3.6 \times 10^{11} $ & 30.86 & -18.16 & 152.63 & 21.46 \\ \hline
16 & 6.41 & 1.16 & $ 1.4 \times 10^{7} $ & $ 3.6 \times 10^{43} $ & $ 4.4 \times 10^{8} $ & $ 1.1 \times 10^{13} $ & 26.87 & -22.15 & \begin{tabular}[c]{@{}c@{}}Not Detectable\end{tabular} & Not Applicable \\ \hline
17 & 6.42 & 2.23 & $ 6.6 \times 10^{5} $ & $ 3.5 \times 10^{41} $ & $ 8.2 \times 10^{7} $ & $ 1.8 \times 10^{12} $ & 28.76 & -20.26 & 153.28 & 21.28 \\ \hline
18 & 6.5 & 0.04 & $ 4.7 \times 10^{7} $ & $ 3.3 \times 10^{43} $ & $ 4.8 \times 10^{8} $ & $ 1.1 \times 10^{13} $ & 26.88 & -22.18 & \begin{tabular}[c]{@{}c@{}}Not Detectable\end{tabular} & Not Applicable \\ \hline
19* & 6.5 & 0.17 & $ 3 \times 10^{6} $ & $ 1.2 \times 10^{42} $ & $ 4.8 \times 10^{8} $ & $ 1 \times 10^{13} $ & 27.29 & -21.78 & 26.34 & 720.46 \\ \hline
20 & 6.51 & 0.67 & $ 1.2 \times 10^{6} $ & $ 6.3 \times 10^{41} $ & $ 8.8 \times 10^{7} $ & $ 2.2 \times 10^{12} $ & 28.91 & -20.16 & 81.70 & 74.91 \\ \hline
21 & 6.51 & 0.79 & $ 1.3 \times 10^{6} $ & $ 7.8 \times 10^{41} $ & $ 2 \times 10^{7} $ & $ 9.1 \times 10^{11} $ & 30.2 & -18.87 & 72.44 & 95.28 \\ \hline
22 & 6.51 & 0.88 & $ 4 \times 10^{6} $ & $ 6.4 \times 10^{42} $ & $ 1.2 \times 10^{8} $ & $ 1.3 \times 10^{12} $ & 28.51 & -20.55 & \begin{tabular}[c]{@{}c@{}}Not Detectable\end{tabular} & Not Applicable \\ \hline
23 & 6.55 & 1.23 & $ 1.5 \times 10^{7} $ & $ 3.5 \times 10^{43} $ & $ 8.5 \times 10^{8} $ & $ 1.6 \times 10^{13} $ & 26.45 & -22.61 & \begin{tabular}[c]{@{}c@{}}Not Detectable\end{tabular} & Not Applicable \\ \hline
24 & 6.6 & 0.31 & $ 6.8 \times 10^{5} $ & $ 1.7 \times 10^{41} $ & $ 6.3 \times 10^{7} $ & $ 1.8 \times 10^{12} $ & 28.9 & -20.20 & 147.28 & 23.05 \\ \hline
25 & 6.61 & 0.28 & $ 1.4 \times 10^{7} $ & $ 6.4 \times 10^{43} $ & $ 5.8 \times 10^{8} $ & $ 1.2 \times 10^{13} $ & 26.33 & -22.77 & \begin{tabular}[c]{@{}c@{}}Not Detectable\end{tabular} & Not Applicable \\ \hline
26 & 6.68 & 0.71 & $ 1.8 \times 10^{8} $ & $ 1.2 \times 10^{44} $ & $ 4.8 \times 10^{8} $ & $ 9.7 \times 10^{12} $ & 25.87 & -23.27 & \begin{tabular}[c]{@{}c@{}}Not Detectable\end{tabular} & Not Applicable \\ \hline
27 & 6.69 & 0.45 & $ 6.8 \times 10^{5} $ & $ 1.1 \times 10^{40} $ & $ 9 \times 10^{6} $ & $ 4.7 \times 10^{11} $ & 30.96 & -18.18 & 145.09 & 23.75 \\ \hline
28 & 6.7 & 0.07 & $ 6.9 \times 10^{5} $ & $ 7.5 \times 10^{39} $ & $ 6.1 \times 10^{6} $ & $ 3.9 \times 10^{11} $ & 31.08 & -18.06 & 144.11 & 24.08 \\ \hline
29 & 6.76 & 1.4 & $ 1 \times 10^{7} $ & $ 4.2 \times 10^{43} $ & $ 4.8 \times 10^{8} $ & $ 8.8 \times 10^{12} $ & 26.79 & -22.39 & \begin{tabular}[c]{@{}c@{}}Not Detectable\end{tabular} & Not Applicable \\ \hline
30 & 6.79 & 0.40 & $ 6.9 \times 10^{5} $ & $ 8 \times 10^{38} $ & $ 2.3 \times 10^{6} $ & $ 1.6 \times 10^{11} $ & 32.42 & -16.76 & 142.15 & 24.74 \\ \hline
31* & 6.8 & 0.27 & $ 1.5 \times 10^{6} $ & $ 2 \times 10^{42} $ & $ 4.6 \times 10^{8} $ & $ 6.8 \times 10^{12} $ & 27.28 & -21.90 & 56.30 & 157.75 \\ \hline
32 & 6.91 & 0.68 & $ 1.5 \times 10^{6} $ & $ 4.5 \times 10^{41} $ & $ 3.3 \times 10^{8} $ & $ 6.1 \times 10^{12} $ & 27.57 & -21.64 & 94.08 & 56.49 \\ \hline
33 & 6.92 & 0.88 & $ 7.1 \times 10^{5} $ & $ 7.3 \times 10^{39} $ & $ 1.4 \times 10^{7} $ & $ 5.7 \times 10^{11} $ & 30.5 & -18.72 & 137.90 & 26.29 \\ \hline
34 & 6.96 & 2.71 & $ 1.4 \times 10^{6} $ & $ 6.3 \times 10^{41} $ & $ 5.7 \times 10^{7} $ & $ 1.3 \times 10^{12} $ & 29.34 & -19.91 & 62.59 & 127.61 \\ \hline
35 & 6.98 & 1.80 & $ 7.8 \times 10^{5} $ & $ 9.6 \times 10^{40} $ & $ 2.3 \times 10^{7} $ & $ 8.4 \times 10^{11} $ & 30.02 & -19.23 & 132.05 & 28.67 \\ \hline
36* & 7.01 & 0.65 & $ 1.1 \times 10^{6} $ & $ 4 \times 10^{41} $ & $ 6.9 \times 10^{7} $ & $ 2.1 \times 10^{12} $ & 29.13 & -20.12 & 96.34 & 53.87 \\ \hline
37 & 7.04 & 2.49 & $ 1.3 \times 10^{7} $ & $ 2 \times 10^{43} $ & $ 3.5 \times 10^{8} $ & $ 9.1 \times 10^{12} $ & 27.52 & -21.73 & \begin{tabular}[c]{@{}c@{}}Not Detectable\end{tabular} & Not Applicable \\ \hline
38 & 7.2 & 0.15 & $ 1.2 \times 10^{7} $ & $ 3.9 \times 10^{43} $ & $ 4.2 \times 10^{8} $ & $ 8.7 \times 10^{12} $ & 27.13 & -22.19 & \begin{tabular}[c]{@{}c@{}}Not Detectable\end{tabular} & Not Applicable \\ \hline
39 & 7.24 & 1.32 & $ 1.2 \times 10^{6} $ & $ 2.2 \times 10^{42} $ & $ 5 \times 10^{7} $ & $ 1.2 \times 10^{12} $ & 29.33 & -19.99 & 84.97 & 69.25 \\ \hline
40 & 7.24 & 2.19 & $ 2.2 \times 10^{6} $ & $ 1.5 \times 10^{42} $ & $ 2 \times 10^{8} $ & $ 2.5 \times 10^{12} $ & 28.25 & -21.07 & 19.22 & 1352.89 \\ \hline
41 & 7.32 & 1.33 & $ 7.3 \times 10^{5} $ & $ 5.4 \times 10^{38} $ & $ 4.4 \times 10^{7} $ & $ 9 \times 10^{11} $ & 29.45 & -19.90 & 128.22 & 30.41 \\ \hline
42 & 7.36 & 1.28 & $ 3.5 \times 10^{6} $ & $ 2.4 \times 10^{42} $ & $ 2.7 \times 10^{8} $ & $ 4.1 \times 10^{12} $ & 28.01 & -21.38 & \begin{tabular}[c]{@{}c@{}}Not Detectable\end{tabular} & Not Applicable \\ \hline
43 & 7.4 & 0.09 & $ 7.4 \times 10^{5} $ & $ 4 \times 10^{40} $ & $ 3 \times 10^{7} $ & $ 8.8 \times 10^{11} $ & 29.57 & -19.82 & 126.26 & 31.36 \\ \hline
44 & 7.48 & 0.61 & $ 7.5 \times 10^{5} $ & $ 4.6 \times 10^{39} $ & $ 4.7 \times 10^{7} $ & $ 1.4 \times 10^{12} $ & 29.31 & -20.12 & 123.98 & 32.53 \\ \hline
45 & 7.48 & 0.72 & $ 7.5 \times 10^{5} $ & $ 3 \times 10^{39} $ & $ 5.7 \times 10^{5} $ & $ 8.1 \times 10^{10} $ & 32.88 & -16.54 & 124.48 & 32.27 \\ \hline
46 & 7.53 & 1.41 & $ 1.8 \times 10^{6} $ & $ 9.8 \times 10^{41} $ & $ 1.4 \times 10^{8} $ & $ 2.1 \times 10^{12} $ & 28.59 & -20.83 & 37.11 & 363.07 \\ \hline
47 & 7.58 & 0.52 & $ 7.6 \times 10^{5} $ & $ 2.9 \times 10^{40} $ & $ 1.1 \times 10^{7} $ & $ 4.4 \times 10^{11} $ & 30.68 & -18.78 & 121.53 & 33.86 \\ \hline
48 & 7.62 & 1.05 & $ 1.1 \times 10^{7} $ & $ 2.3 \times 10^{43} $ & $ 5.3 \times 10^{8} $ & $ 8.2 \times 10^{12} $ & 27.39 & -22.07 & \begin{tabular}[c]{@{}c@{}}Not Detectable\end{tabular} & Not Applicable \\ \hline
49 & 7.75 & 1.18 & $ 7.9 \times 10^{5} $ & $ 3.7 \times 10^{40} $ & $ 2.9 \times 10^{7} $ & $ 8.8 \times 10^{11} $ & 29.86 & -19.66 & 117.33 & 36.32 \\ \hline
50 & 7.9 & 0.19 & $ 3.3 \times 10^{7} $ & $ 4.3 \times 10^{43} $ & $ 5.2 \times 10^{8} $ & $ 7.2 \times 10^{12} $ & 27.2 & -22.36 & \begin{tabular}[c]{@{}c@{}}Not Detectable\end{tabular} & Not Applicable \\ \hline
51 & 7.98 & 0.79 & $ 1.1 \times 10^{6} $ & $ 8.9 \times 10^{40} $ & $ 6 \times 10^{6} $ & $ 3.2 \times 10^{11} $ & 31.41 & -18.18 & 87.25 & 65.69 \\ \hline
52* & 8 & 0.01 & $ 2.7 \times 10^{6} $ & $ 8.8 \times 10^{42} $ & $ 4.1 \times 10^{8} $ & $ 4.1 \times 10^{12} $ & 27.62 & -21.96 & 10.65 & 4408.89 \\ \hline
53 & 8.01 & 0.39 & $ 8.7 \times 10^{5} $ & $ 1.7 \times 10^{40} $ & $ 1.6 \times 10^{7} $ & $ 6.1 \times 10^{11} $ & 30.68 & -18.90 & 108.16 & 42.74 \\ \hline
54 & 8.02 & 0.52 & $ 8 \times 10^{5} $ & $ 1.9 \times 10^{40} $ & $ 1.6 \times 10^{7} $ & $ 6.1 \times 10^{11} $ & 30.3 & -19.28 & 112.29 & 39.66 \\ \hline
55 & 8.51 & 0.29 & $ 1.2 \times 10^{6} $ & $ 1.5 \times 10^{41} $ & $ 2.3 \times 10^{7} $ & $ 7.1 \times 10^{11} $ & 30.16 & -19.58 & 75.95 & 86.69 \\ \hline
56 & 8.99 & 0.07 & $ 9 \times 10^{5} $ & $ 4.3 \times 10^{40} $ & $ 3.2 \times 10^{7} $ & $ 6.3 \times 10^{11} $ & 30.0 & -19.88 & 93.90 & 56.71 \\ \hline
\end{tabular}%
}
\end{table*}

\section{Comparison with non-merging MBH}\label{nonmerg}

\begin{figure*}
        \centering
        
          \includegraphics[scale=0.45]
          {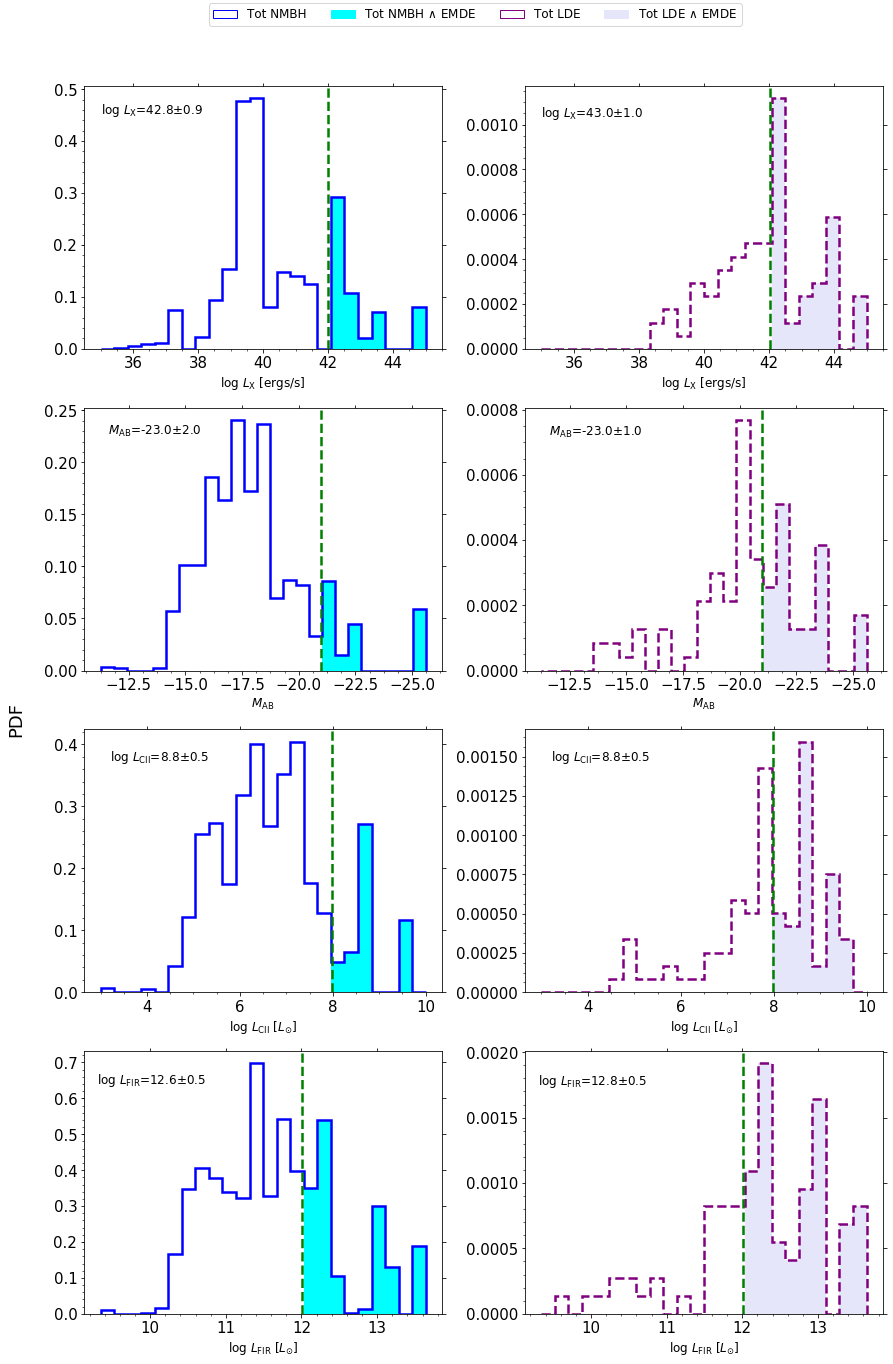}
         
    \caption{PDFs of different observable properties of MBHs associated with a host galaxy which eventually merge within the LISA band (purple dashed) and do not merge referred to as NMBH,  (blue solid) at $z>6$.
    In the left column, the cyan bars represent the non-merging MBHs which are detectable by EM telescopes and in the right column, the lavender bars represent the LDEs which are also EMDEs (see also Fig.~\ref{fig:L_comp}). We report results for different wavelength bands: X-ray (first row), UV (second row), [CII] (third row), FIR (bottom row). In each panel, we also report the average and 1$\sigma$ values of EM bright events.}

    \label{fig:nonmerg}
\end{figure*}

In this section, we compare the observable properties of LDEs with MBHs in the same mass range, seeded in a host galaxy, which are not associated with any mergers (non-merging MBHs, NMBHs). In Fig.~\ref{fig:nonmerg}, we show in the left column, the observable properties of all MBHs associated with a host galaxy which are not involved in any coalescence. Whereas, in the right column we show the observable properties of host galaxies of LDEs. We also show the EM bright events in both cases as described in Sec.~\ref{sec:EMDEs}. We note that in our simulations, only $\sim0.3\%$ of all MBHs eventually result in mergers. From fig.~\ref{fig:nonmerg}, we can see there is no significant difference in the distribution of observable properties of merging MBHs detectable by LISA and non-merging MBHs. Hence, within our simulation limits, we cannot distinguish LDE population from NMBH population based on their observable EM properties.

\label{lastpage}
\end{document}